\documentclass[aps,prc,superscriptaddress,twocolumn]{revtex4}

\usepackage{amssymb}
\usepackage{graphicx}
\usepackage{epsfig}%
\usepackage{bm}
\usepackage{hyperref}
\usepackage{mathrsfs}
\usepackage{multirow}
\usepackage{footmisc}
\usepackage{threeparttable}
\usepackage{booktabs}
\usepackage{appendix}
\usepackage{float}

\begin{document}

\title{In-medium pion dispersion relation  and medium correction of $N\pi\leftrightarrow \Delta$ near the threshold energy of pion production}


\author{Ying Cui}
\email{cuiying@ciae.ac.cn}
\affiliation{China Institute of Atomic Energy, Beijing 102413, China}

\author{Yingxun Zhang}
\email{zhyx@ciae.ac.cn}
\affiliation{China Institute of Atomic Energy, Beijing 102413, China}
\affiliation{Guangxi Key Laboratory Breeding Base of Nuclear Physics and Technology, Guangxi Normal University, Guilin 541004, China}

\author{Zhuxia Li}
\affiliation{China Institute of Atomic Energy, Beijing 102413, China}

\begin{abstract}
Transport models cannot simultaneously explain very recent data on pion multiplicities and pion charged ratios of Sn+Sn in the reaction at 0.27 A GeV. This stimulates further  investigations on the pion dispersion relation, in-medium $N\pi\to \Delta$ cross sections and $\Delta \to N \pi$ decay widths near the threshold energy or at subthreshold energy of pion production in isospin asymmetric nuclear matter. In this study, the pion dispersion relation, in-medium $N\pi\to \Delta$ cross section and $\Delta \to N \pi$ decay width near the threshold energy are investigated in isospin asymmetric nuclear matter by using the  one-boson-exchange model. With the consideration of the energy conservation effect, the in-medium $N\pi\to\Delta$ cross sections are enhanced at $s^{1/2}<1.11$ GeV in  nuclear medium. The prediction of pion multiplicity and $\pi^-/\pi^+$ ratios near the threshold energy can be modified if this effect is considered in  transport model simulations.
\end{abstract}

\date{\today}


\pacs{Valid PACS appear here}

\maketitle

\section{Introduction}
Symmetry energy plays an important role in understanding the isospin asymmetric subject, e.g., in neutron stars ~\cite{Fattoyev2013,Tsang2019} and in observables of neutron rich heavy ion collisions ~\cite{BALi02,LWChen05,Zhang20}. However the density dependence of symmetry energy, especially at high density, is still largely  unknown. In addition to the efforts on the constraints of the symmetry energy at suprasaturation density by analyzing the neutron star merging events~\cite{Abbott2017,Abbott2018,Fattoyev2018,Annala2018,Abbott2019,Zhang2019,Xie2019,Tsang2019,Tsang2019v2,Russotto2011,Pagano2011},  constraints related to heavy ion collisions are also needed.
The ratio of multiplicity of $\pi^-$ to $\pi^+$, known as $\pi^-/\pi^+$ ratios, in heavy ion collisions with neutron rich beam and target at the $\pi$ production threshold energy was supposed to be a sensitive observable to probe the density dependence of the symmetry energy at suprasaturation density~\cite{Sako2014}.

The pion data of Au+Au at beam energy ranging from 0.4 A GeV to 1.2 A GeV~\cite{Resid07} were used for extracting the information of the symmetry energy. However, contradictory conclusions on the symmetry energy were obtained when comparing data to calculations with different transport models~\cite{Xiao2009,Feng2010,Xie2013,Hong2014,Song2015,Cozma2017}. In addition to the model uncertainties which arise from the philosophy of solving the high dimensionality transport equation, another important reason is that the sensitivity of $\pi^-/\pi^+$ ratios may be be strong enough to clearly distinguish the stiffness of symmetry energy at higher beam energy where the nucleon-nucleon collisions play a dominant role instead of the mean field. This stimulated the remeasurement of pion multiplicities and $\pi^-/\pi^+$ ratios near  the $\Delta$ threshold energy  by using  neutron rich reaction systems.

Recently, the MSU group measured the charged pion multiplicities for $^{132,112,108}$Sn+$^{124,112}$Sn by using the S$\pi$RIT Time Projection Chamer at 0.27A GeV, which is subthreshold energy ~\cite{Jhang2021}. The energy spectral of single $\pi^-/\pi^+$ ratios, i.e., $R(\pi^-/\pi^+)=\frac{dM_{\pi^-}}{dE_k}/\frac{dM_{\pi^+}}{dE_k}$, and double ratios $DR(\pi^-/\pi^+)=R_{\mathrm{A}}(\pi^-/\pi^+)/R_{\mathrm{B}}(\pi^-/\pi^+)$ (A is $^{132}$Sn+$^{124}$Sn,  B is $^{108}$Sn+$^{112}$Sn ) will be provided, and they may have a more exclusive sensitivity to the density dependence of the symmetry energy  than the total multiplicity of pions~\cite{Hong2014,Tsang17}.

However, the behavior of the pion energy spectral or pion flow could also be sensitive to the pion potential~\cite{Xiong1993,Buss2012,Guo2015,Feng2016,Liu2018}. The reason is that,  for beam energy below 0.3 A GeV, the pions are mainly produced through the low mass $\Delta$s. The produced pions via low mass $\Delta$s decay have smaller momentum ($|\mathbf{k}|<0.119$ GeV $< m_\pi$), so they have  longer mean-free-path given that the cross sections of $\pi+N$ near the threshold energies are relative small~\cite{Ono2019}. Consequently, one can expect that the in-medium effects on pion propagation and collision gradually become more important, and the on-shell transport seems reasonably accurate. The relativistic Vlasov-Uehling-Uhlenbeck (RVUU) model calculations showed that the $\pi^-/\pi^+$ ratio is reduced by approximately 10\% by considering the in-medium pion dispersion relation, and a large effect is also observed in isospin-dependent Boltzmann-Uehling-Uhlenbeck (IBUU)~\cite{Guo2015} calculations at subthreshold energy. There  exists model dependence on the in-medium effects on the pion production mechanism  in the transport model simulations partly  due to the separate treatments on pion potential, the $\pi N\to \Delta$ cross sections and $\Delta \to \pi N$ decay widths. Thus, providing a theoretical description of the pion potential, $\pi N\to \Delta$ cross sections and $\Delta \to \pi N$ decay widths in isospin asymmetric nuclear matter from the same Lagrange is very useful to achieve a deep understanding of the pion production mechanism and reduce the model uncertainties related to the medium corrections separately for pion potential and $N\pi\to \Delta /\Delta\to N\pi$ .

Generally, the pion potential can be obtained from the phenomenological pion potential~\cite{Buss2012,Guo2015,Feng2016,Cozma2017,Liu2018}, or from effective methods, such as the closed-time path green functions method~\cite{Mao1999prc}or the chiral perturbation theory~\cite{Kaiser,Girlanda2005}. Dmitriev el at. studied on the in-medium pion dispersion relation by the pion self-energy via meson exchange interaction~\cite{Dmitriev1985} for symmetric nuclear matter. Then, a study by Guangjun Mao also discussed the pion dispersion relation with the relativistic form of pion self-energy in symmetric nuclear matter and its effect on the $N\pi\to \Delta$ cross section and $\Delta\to N\pi$ decay width~\cite{Mao1999prc,QingfengLi2017v2}.

This increasing interest on the study of isospin asymmetric nuclear matter led Kaiser el at. to study the pion s-wave self-energy in isospin asymmetric nuclear matter based on the chiral perturbation theory up to the two-loop approximation~\cite{Kaiser}. With the s-wave pion self-energy reported in Ref.~\cite{Kaiser}, Zhen Zhang el at. also added p-wave pion potential for the estimation of the in-medium $N\pi\to \Delta$ cross sections and $\Delta \to \pi N$ decay widths by including N and $\Delta$ masses in free space~\cite{ZhenZhang2017}. Qingfeng Li et al. discussed  the $N\pi\to \Delta$ cross sections and $\Delta \to \pi N$ decay widths in asymmetric nuclear matter based on the closed-time path green function methods~\cite{QingfengLi2017v2}, but the pion dispersion relation they used  was still in symmetric nuclear matter~\cite{Mao1999prc}. In these calculations, the energy conservation is an important issue and should be carefully considered in isospin asymmetric nuclear matter for $NN\to N\Delta$~\cite{Cui2018}.

In this study, we investigated the pion self-energy, the in-medium $N\pi\to \Delta $ cross section and the $\Delta\to N\pi$ decay width in asymmetric nuclear matter with the consideration of energy conservation and effective mass splitting effects based on relativistic form interaction. Given that we focused on the pion self-energy, the medium effects of pion absorption cross section and $\Delta$ decay width,  we did not include the $\Delta$ width in the $\Delta$ propagator for calculating the pion self-energy and other related results with an approximation. The paper is organized as follows. In Sec.~\ref{Approaches}, we introduce the theoretical model on the pion self-energy, $N\pi\to \Delta $ cross section and $\Delta$ decay width. Then, the in-medium pion dispersion relation, $N\pi\to \Delta $ cross section and $\Delta \to \pi N$ decay width are presented and discussed  in Sec.~\ref{Results}. Finally, a summary and conclusion are provided in Sec.~\ref{summary}.

\section{Theoretical Model}
\label{Approaches}
 Based on the particle-hole and $\Delta$-hole  with  relativistic form interaction, we studied  the in-medium pion dispersion relation and calculated the $\pi N\to \Delta$ cross sections and $\Delta \to \pi N $ decay widths in an asymmetric medium.  
The Lagrangian density we adopted is as follows \cite{Cui2018,Cui2019,Huber1994,Machleidt1987,Benmerrouche1989}
\begin{eqnarray}
\label{eq:Lag}
\mathcal{L}=\mathcal{L}_F+\mathcal{L}_I
\end{eqnarray}
where $\mathcal{L}_F$ is the free Lagrangian for the nucleon and $\Delta$~\cite{Cui2018, Cui2019}. The interaction part of the Lagrangian is
\begin{eqnarray}
\label{eq:Lag}
\mathcal{L}_I=\frac{g_{\pi NN}}{m_{\pi}}\bar{\Psi}\gamma_{\mu}\gamma_{5}\bm{\tau} \cdot\Psi\partial^{\mu}\bm{\pi}+\frac{g_{\pi N\Delta}}{m_{\pi}}\bar{\Delta}_{\mu}\bm{\mathcal{T}}\cdot \Psi\partial^{\mu}\bm{\pi}+h.c.
\end{eqnarray}
Here, $\bm{\tau}$ is the isospin matrices of the nucleon~\cite{Machleidt1987}, and $\bm{\mathcal{T}}$ is the isospin transition matrix between the isospin 1/2 and 3/2 fields~\cite{Huber1994}.

The pion dispersion relation in a nuclear medium is
\begin{equation}
\label{eq:pid}
\omega_{\pi^i}^2=m_{\pi^i}^2+\mathbf{k}^2+\Pi(k).
\end{equation}
where $\pi^i$ denotes different isospin states of the pion, i.e., $\pi^+$, $\pi^0$, and $\pi^-$. Here, $\Pi(k)$ is the pion self-energy; it includes the particle-hole ($\Pi_N(k)$) and $\Delta$-hole parts ($\Pi_\Delta(k)$), i.e.,
\begin{equation}
\label{eq:pid1}
\Pi(k)=\Pi_N(k)+\Pi_\Delta(k).
\end{equation}
For the on-shell pion dispersion relation in Eq.~(\ref{eq:pid}), the $\Pi(k)$ should be the real part ($\mathrm{Re}\Pi(k)$). For convenience, all the Re notions in the $\Pi(k)$ were ignored in this study. The lowest order $\pi$ self-energies in nuclear matter are
\begin{eqnarray}
\Pi_{N}&=&(-i)(\frac{g_{\pi NN}}{m_{\pi}})^2 \langle t^\prime |\tau^{\lambda} | t \rangle\langle t |\tau^{\dagger\lambda^{\prime}} | t^\prime \rangle \delta_{\lambda\lambda^{\prime}}\\
&&\times\int\frac{d^4 q}{(2\pi)^4} \mathbf{Tr}[k\!\!\!\!/\gamma_{5} G_{N}(q+k)k\!\!\!\!/\gamma_{5}G_{N}(q)],\nonumber
\end{eqnarray}
\begin{eqnarray}
\Pi_{\Delta}&=&(-i)(\frac{g_{\pi N\Delta}}{m_{\pi}})^2 \langle t^\prime |\mathcal{T}^{\lambda} | t \rangle\langle t |\mathcal{T}^{\dagger\lambda^{\prime}} | t^\prime \rangle \delta_{\lambda \lambda^{\prime}}\\
&&\times\int\frac{d^4q}{(2\pi)^4} \mathbf{Tr}[k_{\mu}k_{\nu}G^{\mu\nu}_{\Delta}(q+k)G_{N}(q)],\nonumber
\end{eqnarray}
where isospin matrix $\tau^{\lambda}$ can be $\tau^{+}$,  $\tau^{0}$ and  $\tau^{-}$ as in Ref.~\cite{Ericson1988}. Here we take the isospin factors $\langle t^\prime |\tau^{\lambda} | t \rangle=I_{NN}$, and $\langle t^\prime |\mathcal{T}^{\lambda} | t \rangle=I_{N\Delta}$, which can be found in appendix~\ref{AppendixA}.

The nucleon and $\Delta$ propagators in a nuclear medium can be expressed using the  above Lagrangian as follows:
\begin{eqnarray} \label{eq:propn}
G_{N}(q_0, \mathbf{q})&&= \frac{q\!\!\!\!/+m_{N}}{q^2_0-E^2_{N}(q)+i\epsilon}\\
&&+i2\pi\frac{q\!\!\!\!/+m_{N}}{2E_{N}(q)}n(|\mathbf{q}|)\delta(q_0-E_{N}(q))\nonumber
\end{eqnarray}
\begin{eqnarray}\label{eq:propd}
G^{\mu\nu}_{\Delta}(q_0, \mathbf{q})&&= \frac{\mathcal{P}^{\mu\nu}}{q^2_0-E^2_{\Delta}(q)+i\epsilon} \\
&&+i2\pi\frac{q\!\!\!\!/+m_{0,\Delta}}{2E_{\Delta}(q)}n(|\mathbf{q}|)\delta(q_0-E_{\Delta}(q))\nonumber
\end{eqnarray}
where $\mathcal{P}^{\mu\nu}$ is
\begin{equation}
\label{eq:dmunu}
\mathcal{P}^{\mu\nu}=-(q\!\!\!\!/+m_{0,\Delta})[g^{\mu\nu}-\frac{1}{3}\gamma^{\mu}\gamma^{\nu}-\frac{2q^{\mu}q^{\nu}}{3m^{2}_{0,\Delta}}+\frac{q^{\mu}\gamma^{\nu}-q^{\nu}\gamma^{\mu}}{3m_{0,\Delta}}].
\end{equation}
Here, $q\!\!\!\!/=q^{\mu}\gamma_{\mu}$ and $m_{0,\Delta}$ is the pole mass of $\Delta$. The first parts of Eq.~(\ref{eq:propn}) - (\ref{eq:propd})  are the vacuum propagators, and $n(|\mathbf{q}|)$
denotes the  occupation number in the medium. Note that we made an approximation on the $\Delta$ propagator, i.e. we replaced the imaginary part  $i\sqrt{p^2}\Gamma_\Delta(p^2)$ from the Eq.(8) in Ref.\cite{Larionov2002}  with $i\epsilon$ in  Eq.(\ref{eq:propd}) . This is because we wanted to focus on  the in-medium effect on pion self-energy, absorption and production cross sections at a beam energy $<$ 0.4A GeV. If we do not consider the Fermi motion, the estimated maximum $\Delta$ mass is approximately less than 1.13 GeV and the momentum of pion is less than 0.119 GeV/c in $\Delta$'s rest frame. Correspondingly, the decay width of $\Delta$ is also small, below 0.03 GeV. Thus, in the following discussion, we mainly discussed the results of pion self-energy, $N \pi\to \Delta$ cross sections and $\Delta$ decay width within $s^{1/2}_{N\pi}<1.15$ GeV. Concerning the medium effects on pion related issues at a beam energy around and above 1A GeV for heavy ion collisions, $i\sqrt{p^2}\Gamma_\Delta(p^2)$ should be adopted in $\Delta$ propagator in the calculations of pion self-energy, cross section and decay width as in Refs.~\cite{Xiong1993,Mao1999prc,ZhenZhang2017,Friedman1981,Mull1992}.

The pion self-energies can also  be expressed in terms of an analog of the susceptibility $\chi$,
\begin{eqnarray}
\Pi_N=k^2\chi_N\\
\Pi_\Delta=k^2\chi_\Delta
\end{eqnarray}
and the short range correlations is incorporated into $\chi$ 
as follows;
\begin{eqnarray}
\chi_{N}\to \frac{1+(g^{\prime}_{N\Delta}-g^{\prime}_{\Delta\Delta})\chi_{\Delta}}{(1-g^{\prime}_{\Delta\Delta}\chi_{\Delta})(1-g^{\prime}_{NN}\chi_{N})-g^{\prime}_{N\Delta}\chi_{\Delta}g^{\prime}_{N\Delta}\chi_{N}} \chi_{N},\nonumber\\
\chi_{\Delta}\to \frac{1+(g^{\prime}_{N\Delta}-g^{\prime}_{NN})\chi_{N}}{(1-g^{\prime}_{\Delta\Delta}\chi_{\Delta})(1-g^{\prime}_{NN}\chi_{N})-g^{\prime}_{N\Delta}\chi_{\Delta}g^{\prime}_{N\Delta}\chi_{N}} \chi_{\Delta}.\nonumber
\end{eqnarray}
The Migdal parameters for the short-range interaction are $g^{\prime}_{NN}=0.9$ and $g^{\prime}_{N\Delta}=g^{\prime}_{\Delta\Delta}=0.6$ as in Ref.~\cite{Xia1988}.
The detailed calculation and self-energies of $\pi^+$, $\pi^0$ and $\pi^-$ are shown in Appendices~\ref{AppendixB} - \ref{AppendixC}.

For the calculation of the in-medium $\Delta\to N\pi$ decay widths and $\pi N\to \Delta $ cross sections, we used the quasiparticle approximation~\cite{Baym1976} by replacing $m_i \to m^*_i$ and $p_i \to p^*_i$ ($i$ could be nucleon or $\Delta$) in their formula.
The effective momentum can be written as $\textbf{p}_i^*=\textbf{p}_i$ given that  the spatial components of the vector field vanish in the rest nuclear matter, i.e., $\mathbf{\Sigma}=0$. Thus, in the mean field approach, the effective energy reads
\begin{equation}
p_i^{*0}=p^{0}_{i}-\Sigma^{0}_{i}.
\end{equation}
The Dirac effective mass of nucleon and the effective pole mass of $\Delta$ read
\begin{equation}
m^{*}_{i}=m_{i}+\Sigma^{S}_{i}.
\label{eq:efmnd}
\end{equation}
Here $i=n$, $p$, $\Delta^{++}$, $\Delta^{+}$, $\Delta^{0}$ and $\Delta^{-}$.
The details about $\Sigma^{0}_{i}$ and $\Sigma^{S}_{i}$ can be found in Refs.~\cite{Cui2018,Cui2019}. Likewise, the parameters of the relativistic  mean field are the NL$\rho \delta$ as in Ref.~\cite{Liu2002}.

Based on the approximation we adopted, $\mathbf{p}_{N}+\mathbf{k}=\mathbf{p}^*_{N}+\mathbf{k}^*$. The energy conservation is given by the canonical momentum conservation relation, i.e., $E_{\Delta}=E_{N}+\omega$, with $E_{\Delta}=E^*_{\Delta}+\Sigma^{0}_{\Delta}$ and $E_{N}=E^*_{N}+\Sigma^{0}_{N}$,
\begin{eqnarray}
\label{eqmpole}
m^*_{\Delta}+\Sigma^{0}_{\Delta}=E^{*}_{N}+\Sigma^{0}_{N}+\omega(\mathbf{k}).
\end{eqnarray}
By using the effective momenta and masses, one can obtain the in-medium $\Delta\to N\pi$ decay widths and $\pi N\to \Delta$ cross sections. In this study, the in-medium decay widths of a given charged state of $\Delta$, i.e., $\Delta\to N\pi$, are expressed as follows:
\begin{eqnarray}
\Gamma^*&=&\frac{1}{2m^*_{\Delta}}\int \frac{d^3\mathbf{p}^*_{N}}{(2\pi)^32E^*_{N}}\frac{d^3\mathbf{p}^*_{\pi} d\omega}{(2\pi)^3 }\overline{|\mathcal{M}_{\Delta\to N\pi}|^2}\nonumber\\
&&\times\delta(\omega^2-\mathbf{p}^{*2}_{\pi}-m^2_{\pi}-\Pi) \nonumber\\
&&\times(2\pi)^4\delta^3(\mathbf{p}^*_{N}+\mathbf{p}^*_{\pi})\delta(E^*_{N}+\omega+\Delta\Sigma-m^*_{\Delta})\nonumber\\
&=&Z_B\frac{\mathbf{k}^2}{8\pi m^*_{\Delta} E^*_{N}\omega}\frac{\overline{|\mathcal{M}^*_{\Delta\to N\pi}|^2}}{\mid\frac{\mathbf{k}}{E^*_{N}}+\frac{\mathbf{k}}{\omega}\mid},\label{eqdecaypionma}
\end{eqnarray}
where $\mathbf{p}^*_{\pi}=\mathbf{p}_{\pi}=\mathbf{k}$ in the static nuclear medium, and $\Delta\Sigma=\Sigma^0_N-\Sigma^0_\Delta$. The spreading width of $\Delta$ \cite{Hirata1979,Oset1987,Rapp1994,Kim1997,Larionov2003}  from the $\Delta$ absorption and the rescattering processes were neglected in this study, because the  process of $\Delta N\to NN$ and  the multiplicity of pions are relatively scarce. For example, the pion multiplicity  is less than 1 per event for Sn+Sn at 0.27 A GeV and less than 6 for Au+Au at 0.4 A GeV~\cite{Resid07}. $Z_B$ is the wave function renormalization
factor,
\begin{eqnarray}
\label{wavefactor}
Z_B=\frac{1}{1-\frac{1}{2\omega}\frac{\partial \Pi(\omega,\mathbf{k})}{\partial \omega}|_{\omega=E_{\pi}^*}},
\end{eqnarray}
where
\begin{eqnarray}
 E_{\pi}^*=\sqrt{\mathbf{k}^{2}+m^2_{\pi}+\Pi(\omega,\mathbf{k})}.
\end{eqnarray}
Given that we focused on the cross sections used  in the heavy ion collisions near the threshold energy where the low mass $\Delta$s dominate, the pion branch plays the main role for $\Delta$ and the $\Delta$-hole branch  is ignored. Thus, we set $Z_B=1$  as in Ref.~\cite{Mao1999prc}

The in-medium $\pi N\to\Delta$ cross section is expressed as
\begin{eqnarray}
\sigma^{*}_{\pi N\to \Delta}=\frac{\pi f^*(m^*_{\Delta})}{4m^*_{\Delta}}\frac{\overline{|\mathcal{M}^*_{\pi N\to \Delta}|^2}}{|\mathbf{k}|(E^*_{N}+\omega )}\label{eqcrosspionm}.
\end{eqnarray}
where $\overline{|\mathcal{M}^*_{\pi N\to \Delta}|^2}=2\overline{|\mathcal{M}^*_{\Delta \to N \pi}|^2}$.  With the effective mass, the  mass distribution $f^*$ is written as
\begin{equation}
\label{eq:btm}
f^*=\frac{2}{\pi}\frac{m^{*2}_{0,\Delta}\Gamma^*_t}{(m^{*2}_{0,\Delta}-m^{*2}_{\Delta})^2+m^{*2}_{0,\Delta}\Gamma^{*2}_t }.
\end{equation}
where $\Gamma^{*}_t$ is the total decay width of $\Delta \to N \pi$ \cite{Larionov2003}.
Note that $m^*_{\Delta}$ in Eq.~(\ref{eqdecaypionma}),~(\ref{eqcrosspionm}) and~(\ref{eq:btm}) is the energy of the $N\pi$ system in medium and is calculated based on the energy conservation relationship in Eq.~(\ref{eqmpole}) with given $\mathbf{k}$,  which corresponds to the center-of-mass of energy as $\sqrt{s}=m_\Delta=\sqrt{m^2_\pi+\mathbf{k}^2}+\sqrt{m^2_N+\mathbf{k}^2}$ in a $\Delta$ static frame for free space.  With $|\mathbf{k}|=\frac{1}{2\sqrt{s}}\sqrt{(s-(m_\pi+m_N)^2)(s-(m_\pi-m_N)^2)}$, the $\Delta$ effective mass can be $m^*_{\Delta}\equiv m^*_{\Delta}(|\mathbf{k}|)= m^*_{\Delta}(\sqrt{s})$, with $f^*=f^*(\sqrt{s})$, and the in-medium $\pi N\to\Delta$ cross section can be written as
\begin{eqnarray}
\sigma^{*}_{\pi N\to \Delta}=\frac{2\pi^2f^*(\sqrt{s})\Gamma^*}{\mathbf{k}^2}\label{eqcrosspionmv2}.
\end{eqnarray}
The  $\Delta$ effective pole mass  $m^{*}_{0,\Delta}$ above is derived from Eq.~(\ref{eq:efmnd}).

The coupling constants we used were determined by fitting the cross section of $\pi^{+}p\to\Delta^{++}$ and $\Delta\to N\pi$ decay width in free space \cite{pdg2018}. The decay width of $\Delta\to N\pi$ in free space is
\begin{eqnarray}
\label{eqdecaypion}
\Gamma&=&\frac{1}{2m_{\Delta}}\int \frac{d^3\mathbf{p}_{N}}{(2\pi)^32E_{N}}\frac{d^3\mathbf{p}_{\pi} }{(2\pi)^3 2\omega}\overline{|\mathcal{M}_{\Delta\to N\pi}|^2}\nonumber\\
&&\times (2\pi)^4\delta^3(\mathbf{p}_{N}+\mathbf{p}_{\pi})\delta(E_{N}+\omega-m_{\Delta})\nonumber\\
&=&\frac{\mathbf{k}^2}{8\pi m_{\Delta} E_{N}\omega}\frac{\overline{|\mathcal{M}_{\Delta\to N\pi}|^2}}{\mid\frac{\mathbf{k}}{E_{N}}+\frac{\mathbf{k}}{\omega}\mid}
\end{eqnarray}
where $\mathbf{p}_{\pi}=\mathbf{k}$.
Here, $\overline{|\mathcal{M}_{\Delta \to N \pi}|^2}$  is calculated as follows:
\begin{eqnarray}
&&\overline{|\mathcal{M}_{\Delta \to N \pi}|^2}\nonumber\\
&&=\frac{1}{4}\sum_{s_{\Delta}}|\mathcal{M}_{\Delta \to N \pi}|^2\nonumber\\
&&=\frac{g_{\pi N \Delta}^{2}I^{2}_{N\Delta}}{4m^{2}_{\pi}}\sum_{s}\Psi(p_{N})\bar{\Psi}(p_{N})k^{\mu}\Delta_{\mu}(p_{\Delta})\bar{\Delta}_{\nu}(p_{\Delta})k^{\nu}\nonumber\\
&&=\frac{g_{\pi N \Delta}^{2}I^{2}_{N\Delta}}{4m^{2}_{\pi}}Tr[(p\!\!\!\!/_{N}+m_{N})k^{\mu}\mathcal{P}_{\mu\nu}(p_{\Delta})k^{\nu}]\nonumber\\
&&=\frac{2g_{\pi N \Delta}^{2}I^{2}_{N\Delta}}{3m^{2}_{\pi}} (m_{N}+E_{N})\mathbf{k}^2m_{\Delta}\label{Mpion}
\end{eqnarray}

The expression of the $\pi N\to \Delta$ cross section is
\begin{eqnarray}
\sigma_{\pi N\to \Delta}&=&\int dm_{\Delta}f(m_{\Delta})\int \frac{d^3\mathbf{p}_{\Delta}}{(2\pi)^32E_{\Delta}}\frac{\overline{|\mathcal{M}_{\pi N\to \Delta}|^2}}{4E_{N}\omega|\frac{\mathbf{p}_{N}}{E_{N}}-\frac{\mathbf{k}}{\omega}|}\nonumber\\
&\times & (2\pi)^4\delta^3(\mathbf{p}_{N}+\mathbf{k}-\mathbf{p}_{\Delta})\delta(E_{N}+\omega-E_{\Delta}).
\end{eqnarray}
In the $\Delta$ rest frame, $\sigma_{\pi N\to \Delta}$ can be written as
\begin{eqnarray}
\sigma_{\pi N\to \Delta}&=&\frac{\pi f(m_{\Delta})}{4m_{\Delta}E_{N}\omega}\frac{\overline{|\mathcal{M}_{\pi N\to \Delta}|^2}}{|\frac{\mathbf{k}}{E_{N}}+\frac{\mathbf{k}}{\omega}|}.\nonumber\\
\end{eqnarray}
  Note that $f(m_{\Delta})$ is the mass distribution of $\Delta$ resonance in free space, which reads
\begin{equation}
\label{eq:bt}
f(m_{\Delta})=\frac{2}{\pi}\frac{m^{2}_{0,\Delta}\Gamma_t}{(m^{2}_{0,\Delta}-m^{2}_{\Delta})^2+m^{2}_{0,\Delta}\Gamma^2_t }.
\end{equation}

\begin{figure}[htbp]
\begin{center}
    \includegraphics[scale=0.25]{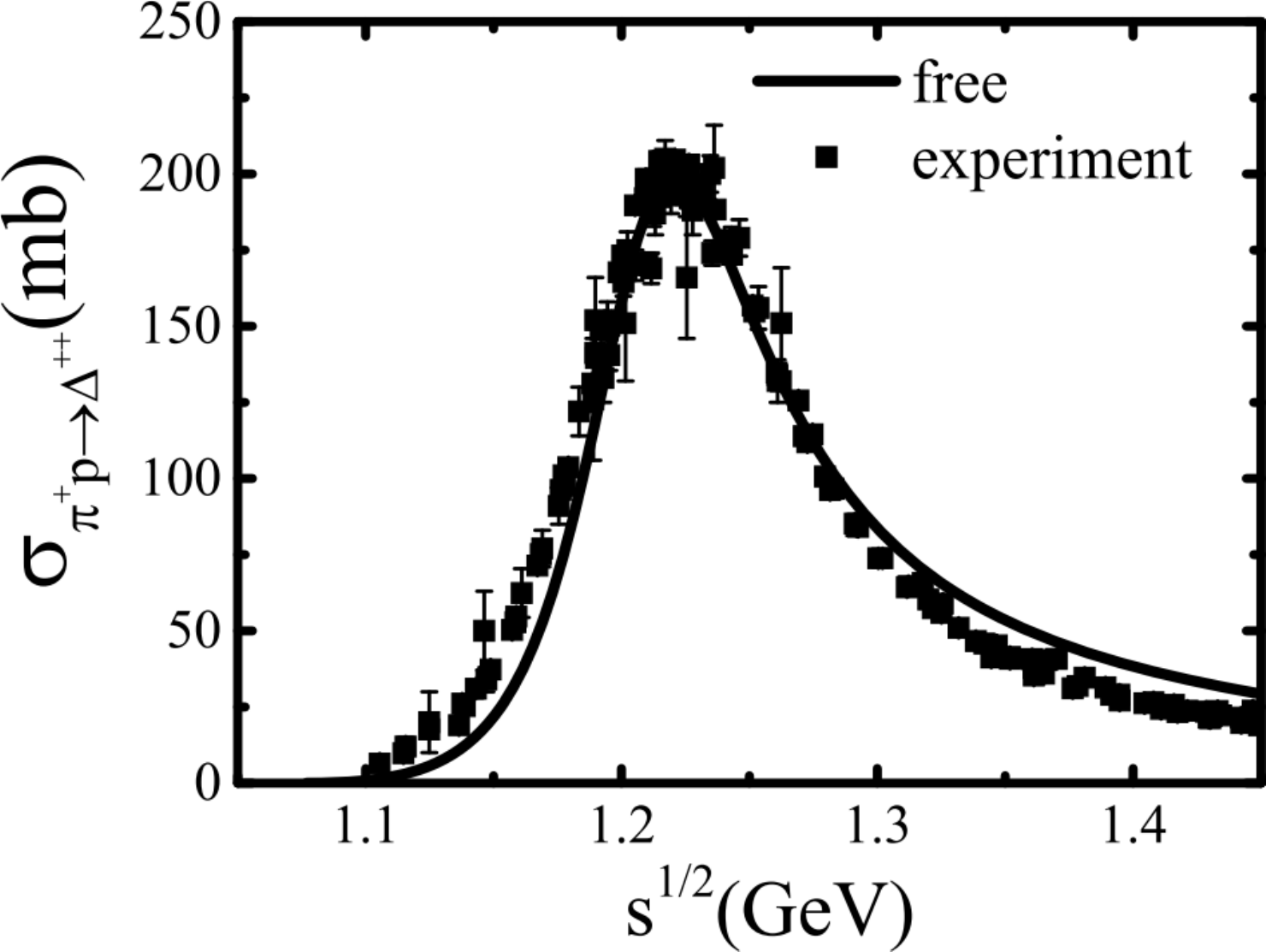}
    \caption{(Color online) $\sigma_{\pi^+ p\rightarrow \Delta^{++}}$ as a function of $s^{1/2}$ in free space; the experimental data are extracted from Ref.~\cite{pdg2018}.}\label{figexp}
\end{center}
\end{figure}

The coupling constants used in this study, i.e., $g_{\pi NN}=1.008$, $g_{\pi N\Delta}=2.3$, and the cut-off $\Lambda^2=exp(-2\mathbf{k}^2/b^2)$ with $b=7m_{\pi}$, are obtained by fitting the experimental data of cross section and decay width~\cite{pdg2018} shown in  Fig.~\ref{figexp}. The decay width $\Gamma=0.120$ GeV for the pole mass $m_{\Delta}=1.232$ GeV can be calculated from Eq.~\ref{eqdecaypion}.

\section{Results and discussions}
\label{Results}
\subsection{Pion dispersion relation}
According to Eq.~(\ref{eq:pid}), the pion dispersion relation has two solutions in , the lower one near the free pion energy $\omega_{F}=\sqrt{m_{\pi}^2 +\mathbf{k}^2}$ , known as the particle-hole branch and higher one near $\omega_{\Delta}=\frac{\mathbf{k}^2}{2m_{\Delta}}+m_{\Delta}-m_{N}$, known as the $\Delta$-hole branch\cite{ZhenZhang2017,Mao1999prc}.
 In Ref.~\cite{ZhenZhang2017}, it was pointed out that the threshold for the $\Delta$ resonance to decay into a pion through the $\Delta$-hole branch is larger than 1.36 GeV. It implies that the $\Delta$ decay into pion via $\Delta$-hole branch is less important given that we focused on the effects near the $\pi$ production threshold energy. Thus, we neglected the decay of $\Delta$ into a pion through the $\Delta$-hole branch as in Ref.~\cite{ZhenZhang2017}.

\begin{figure}[htbp]
\begin{center}
\includegraphics[scale=0.31]{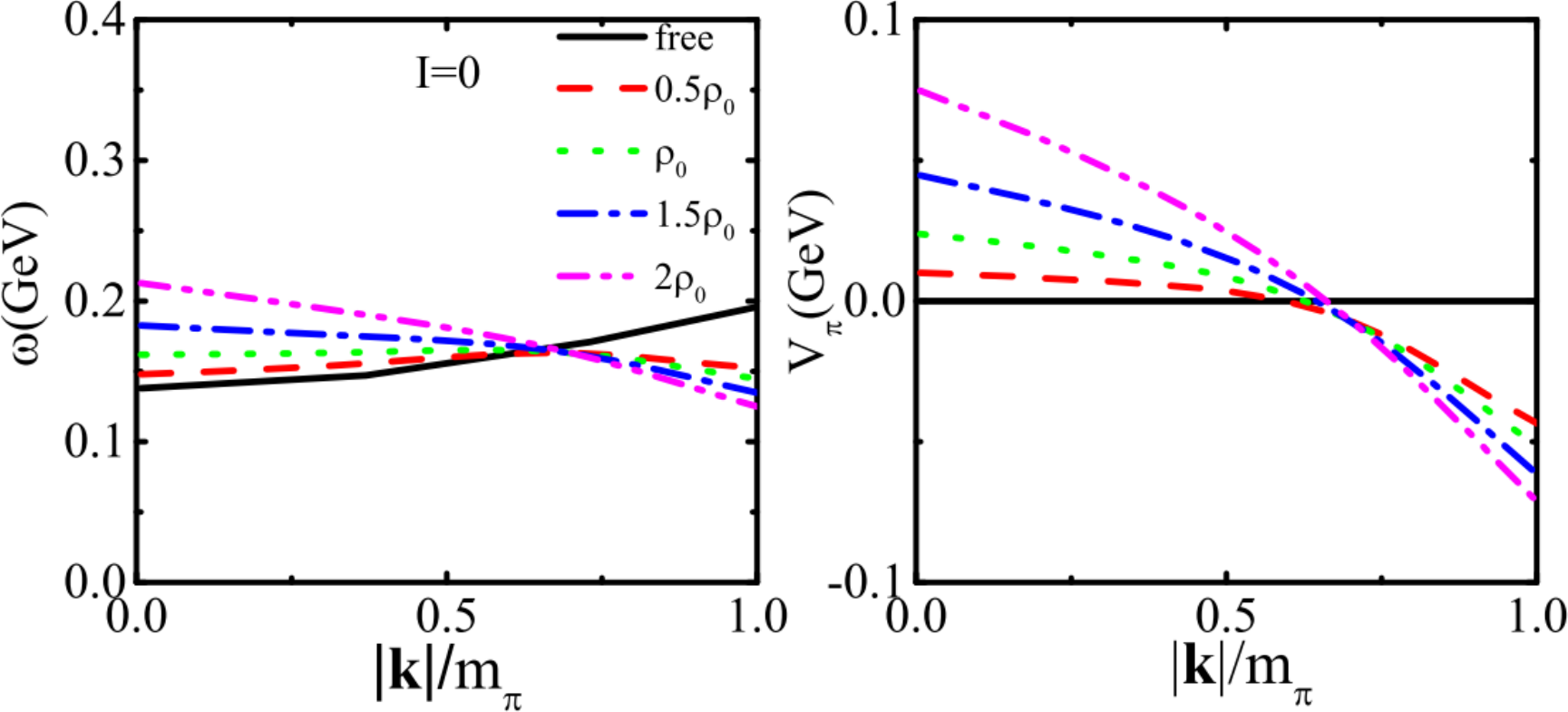}
\caption{(Color online) Left panel: pion dispersion relation at different densities ($0.5\rho_0$, $\rho_0$,  $1.5\rho_0$ and  $2\rho_0$) in symmetric nuclear matter. Right panel: pion optical potential at different densities in symmetric nuclear matter. }\label{fig3v2}
\end{center}
\end{figure}

In Fig.~\ref{fig3v2}, we present the pion dispersion relation $\omega(k)$ and optical potential $V_\pi(k)$ at different densities in symmetric nuclear matter for $|\mathbf{k}|< m_{\pi}$. The black solid, red dashed, green dotted, blue dash-dotted and magenta dash-dotted lines represent $\omega$ in free space, $0.5\rho_0$, $\rho_0$, $1.5\rho_0$ and $2\rho_0$ respectively. The pion optical potential can be written as
\begin{equation}
V_{\pi^i}=\omega_{\pi^i}(\mathbf{k})-\sqrt{m^2_{\pi^i}+\mathbf{k}^2}.
\end{equation}
Note from  Fig.~\ref{fig3v2} that  both the in-medium pion energy $\omega$ and pion optical potential $V_{\pi}$ do not vanish at $|\mathbf{k}|=0$. This is because $k^2$ and $(pk)^2-m^2_{N}k^2$ appear in the relativistic forms of $\Pi_{N}$ and $\Pi_{\Delta}$ (see appendix \ref{AppendixB}).  One can also find that the in-medium pion self-energy do not vanish at momentum $|\mathbf{k}|=0$, i.e., the color lines deviate from the black line at $|\mathbf{k}|$=0, which is different from the results in the nonrelativistic form of pion self-energy reported in Ref.~\cite{ZhenZhang2017}. The in-medium pion energy $\omega$ increases with density increasing at lower momentum  ($|\mathbf{k}|/m_{\pi}<0.66$) while decreases with density at higher momentum ($|\mathbf{k}|/m_{\pi}>0.66$).  The  pion energy in this study is similar to that reported in Refs.~\cite{Dmitriev1985,Mao1999prc}, and the results are the same as  those in Ref.~\cite{Xia1988} according to the nonrelativistic calculation. In the right panel of Fig.~\ref{fig3v2}, we present the pion optical potential for symmetric nuclear matter. The calculation results show that the pions with $|\mathbf{k}|/m_{\pi}<0.66$ experience a repulsive force, while the pions with $ |\mathbf{k}|/m_{\pi}>0.66$ experience an attractive force.
Consequently, one can expect that the pion energy obtained in the heavy ion collisions may show their maximum values at a certain kinetic energy compared to the calculations without considering such the pion potential. The energy slope of pion energy may be a probe to investigate the pion optical potential. For the convenient application of the in-medium pion energy in estimation $\Delta$ decay width and optical potential in transport models, we provide the parameterization form of $\omega$ as a function of $|\mathbf{k}|$ in appendix~\ref{AppendixD}, where $|\mathbf{k}|$ is expanded to $m_{\pi}$.

\begin{figure}[htbp]
\begin{center}
\includegraphics[scale=0.48]{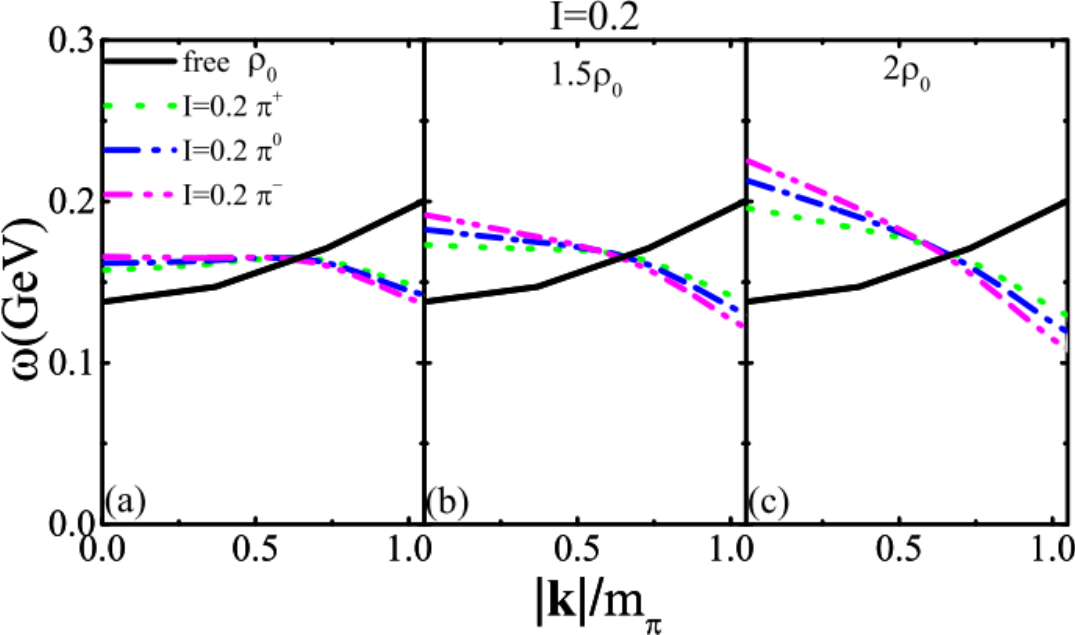}
\caption{(Color online) Pion dispersion relation at different densities ( $\rho_0$, $1.5\rho_0$ and  $2\rho_0$) in asymmetric nuclear matter with $I=0.2$.}\label{fig3v3}
\end{center}
\end{figure}

In Fig.~\ref{fig3v3}, we present the pion dispersion relation in asymmetric nuclear matter for $I=0.2$, where $I=\frac{\rho_n-\rho_p}{\rho_n+\rho_p}$ is the isospin asymmetry. As shown in Fig.~\ref{fig3v3}, $\omega$ is split for different charged state of pions and the difference between $\omega(\pi^-)$ and $\omega(\pi^+)$ is related to the difference between the densities of the neutron and proton, $\rho_n-\rho_p$, in asymmetric nuclear matter.
Interestingly, one can say that $\omega(\pi^-)>\omega(\pi^+)$ at $|\mathbf{k}|<0.66 m_\pi$ and it turns over at $|\mathbf{k}|>0.66 m_\pi$. This behavior agrees with the prediction from nonrelativisitc calculation in Ref.~\cite{ZhenZhang2017} where s-wave plus p-wave potential was adopted. Meanwhile, the magnitude of the splitting of in-medium energy $\omega$, i.e., $|\omega^{-}-\omega^+|$ increases with increasing density in the energy region we discuss.

\begin{figure}[htbp]
\begin{center}
\includegraphics[scale=0.28]{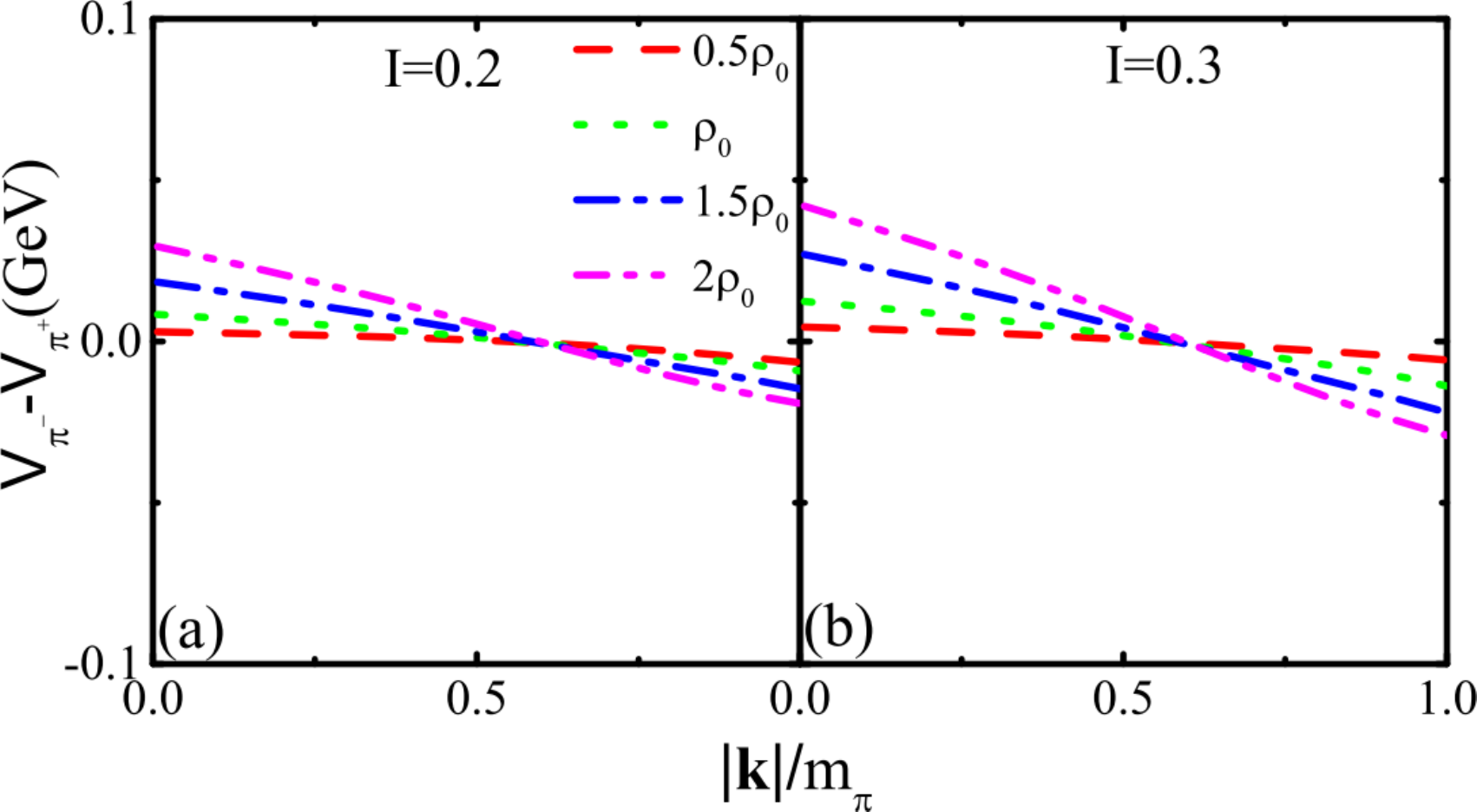}
\caption{(Color online) The  left and right panels are  $V_{\pi^-}-V_{\pi^+}$ at different densities for $I=0.2$ and $I=0.3$ respectively.}\label{fig3v4}
\end{center}
\end{figure}

Fig.~\ref{fig3v4} clearly  depicts that the  $\delta V_{\pi}=V_{\pi^-}-V_{\pi^+}$  depends on the isospin asymmetry, i.e., the amplitude of the  charged pion potential splitting  increases with  increasing isospin asymmetry.

\subsection{In-medium  $\pi N\to \Delta$ cross sections and  $\Delta \to \pi N$ decay widths}

Energy conservation is an important issue to be carefully addressed in the calculations of $N\pi\to \Delta$ and $\Delta\to N\pi$ in isospin asymmetric nuclear matter as well as for $NN\to N\Delta$~\cite{Cui2018}, based on the formulas of the in-medium $\Delta\to N\pi$ decay widths and $N\pi\to\Delta$ cross sections, i.e., in Eqs.~(\ref{eqdecaypionma}) and (\ref{eqcrosspionm}). The $\Delta$ pole mass $m^*_{0,\Delta}$ and distribution function $f^*$ are crucial variables and key parts besides  $\overline{|\mathcal{M}^*_{\pi N\to \Delta}|^2}$,  because $m^*_{0,\Delta}$ can determine the height and position $ (\sqrt{s})$ of the peak of $f^*$. Thus, we first analyzed the values of $m^*_{0,\Delta}$ under the condition of energy conservation, as in Eq.~(\ref{eqmpole}).

\begin{figure}[htbp]
\begin{center}
\includegraphics[scale=0.42]{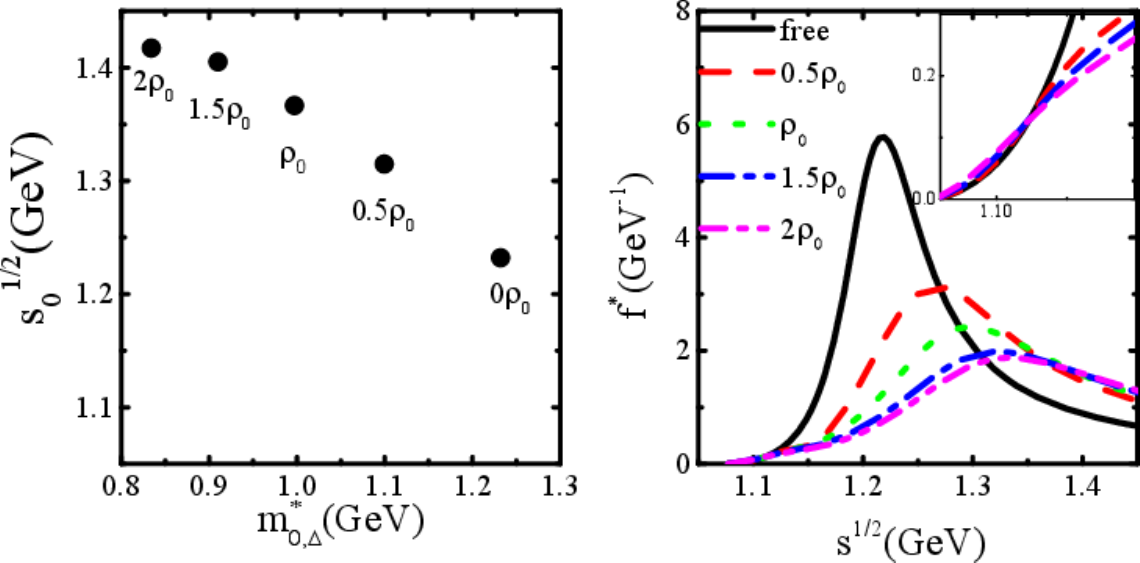}
\caption{(Color online) Left panel: $s^{1/2}_0$ as a function  of $ m^*_{0,\Delta}$ ( details about $s_0^{1/2}$ are in the text).  Right panel: $f^*$ as a function of $s^{1/2}$.}\label{fig3v6}
\end{center}
\end{figure}

The peak of $f^*$ should be around $m^*_{0,\Delta}$, which correspond to a certain momentum $|\mathbf{k}|_{0}$, or energy $s^{1/2}_0=\sqrt{m^2_\pi+\mathbf{k}^2_{0}}+\sqrt{m^2_N+\mathbf{k}^2_{0}}$, and it  satisfies the following relationship:
\begin{equation}
m^*_{0,\Delta}+\Sigma^{0}_{\Delta}=\sqrt{m^*_N+|\mathbf{k}|^{2}_{0}}+\Sigma^{0}_{N}+\omega(|\mathbf{k}|_{0})\label{eqpeak}.
\end{equation}
As shown in Fig.~\ref{fig3v2}, $\omega(\mathbf k)$ at normal density is reduced by 50 MeV  at $|\mathbf{k}|\sim m_{\pi}$. Both $m^*_{0,\Delta}$ and $m^*_N$ decrease with the increase in density, and $m^*_{0,\Delta}-m^*_N=m_{0,\Delta}-m_N$ with $|\Sigma^{0}_{\Delta}-\Sigma^{0}_{N}|=0$ in symmetric nuclear matter. In isospin asymmetric nuclear matter with $I= 0.2$, $m^*_{0,\Delta}-m^*_N$ and  $|\Sigma^{0}_{\Delta}-\Sigma^{0}_{N}|$ are approximately  $40-50$ MeV near $2\rho_0$. It indicates that the values of $\omega$ are key quantities for determining the solution of $|\mathbf k|_0$ from Eq.~(\ref{eqpeak}). When the reductions of $\omega$ is taken into account  and the energy conservation relationship is considered, a larger $|\mathbf{k}|_0$ is expected. Consequently, the position of the peak of $f^*$ moves to a higher energy with smaller effective mass. In the left panel of Fig.~\ref{fig3v6}, we present $s^{1/2}_0$ as a function of $m^*_{0,\Delta}$. It clearly illustrates that the behavior of $s^{1/2}_0$ increases with the decrease in effective mass, or equivalently with the increase in density.

 To obtain a general impression on the aforementioned effects, $f^*$ as a function of $s^{1/2} $ at different densities is presented in symmetric nuclear matter in the right panel of Fig.~\ref{fig3v6}. For a low energy near $s^{1/2}<1.15$ GeV which corresponds to the threshold energy of pion production,  our results show that $f^*$  is enhanced with respect to that in free space at $s^{1/2}<1.11$ GeV (as shown in the inserted panel in Fig.~\ref{fig3v6}) and decreases with respect  relative to $f$ in free space at $s^{1/2}>1.11$ GeV. In addition, the peak of $f^*$ shifts to a higher momentum, which results in the reduction of in-medium $\pi N \to \Delta$ cross sections at high energies. Note that, at $s^{1/2}>$1.3 GeV, there is an enhancement of $f^*$ with respect to the $f$ in free space. The conclusions about this energy region could change by considering the medium effects on the decay width of $\Delta$ and on $\Delta$ propagation.

\begin{figure}[htbp]
\begin{center}
\includegraphics[scale=0.3]{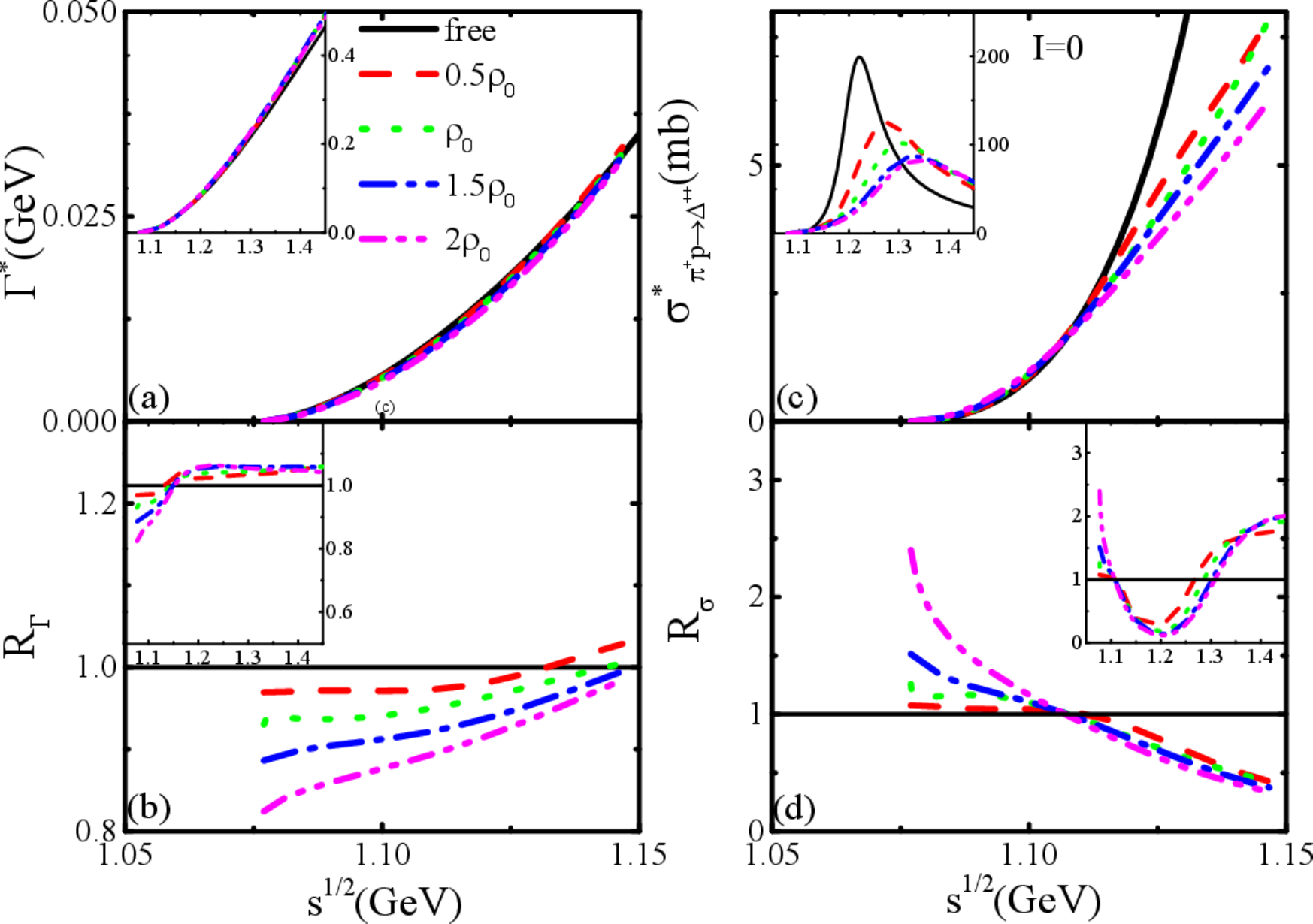}
\caption{(Color online) Left panels: (a) in-medium decay width of $\Delta \to \pi N$ at different densities in symmetric nuclear matter;(b) the medium correction factor $\Gamma^*/\Gamma^{\mathrm{free}}$. Right panels: (c) in-medium cross sections of $\pi^+ p\to \Delta^{++}$ at different densities in symmetric nuclear matter;(d) medium correction factor $\sigma^*/\sigma^{\mathrm{free}}$.}\label{fig3v5}
\end{center}
\end{figure}

Based on the in-medium pion energy and effective masses of N and $\Delta$, the in-medium $\Delta \to \pi N$ decay width and cross section of $\pi^+ p\to \Delta^{++}$ at different densities in symmetric nuclear matter were calculated according to Eq.~(\ref{eqdecaypionma}) and (\ref{eqcrosspionm}); they present in Fig.~\ref{fig3v5} (a) and (c). Panels (b) and (d) are the medium correction factors $R_{\Gamma}=\Gamma^*/\Gamma^{\mathrm{free}}$ and $R_{\sigma}=\sigma^*/\sigma^{\mathrm{free}}$, respectively. As shown in panel (b), the in-medium $\Delta \to \pi N$ decay widths are reduced with respect to that in free space at $s^{1/2}<1.15$ GeV.    The in-medium effect at the energy region ($s^{1/2}<1.15$ GeV) we studied is more evident than that at higher energies ($s^{1/2}>1.15$ GeV). Generally speaking, the medium effect on the decay width is weak,  because the impacts of $m^*_{\Delta}(E^*_{N}+\omega )$ and $\overline{|\mathcal{M}^*_{\pi N\to \Delta}|^2}$ in Eq.~(\ref{eqdecaypion}) conceal each other at higher energies.

Concerning the in-medium pion absorption cross sections $\pi N\to \Delta$, as shown in panel (c) of Fig.~\ref{fig3v5}, they are enhanced at $s^{1/2}<1.11$ GeV ($E_{\mathrm{beam}}\sim$ 0.36 A GeV) and then suppressed at approximately $s^{1/2}>1.11$ GeV  as presented in panel (d). The enhancement of the in-medium pion absorption cross sections $\pi N\to \Delta$ near $s^{1/2}<1.11$ GeV  could lead to the reduction  of pions in the HIC near the threshold energy, while the enhancement of pion production may occur at  $s^{1/2}>1.11$ GeV. 
At $s^{1/2}>1.3$ GeV, the cross sections are enhanced again, but  our results are obtained by neglecting the width of $\Delta$ in the $\Delta$ propagator in the calculation of the pion self-energy. This should be carefully investigated at high energy. Our predictions on the in-medium effects for the $\pi N\to \Delta$ cross sections are similar to the conclusion in Ref.~\cite{ZhenZhang2017}, in which enhancement of in-medium $N\pi\to\Delta$ cross sections near the threshold energies was reported. However, the amplitude of the enhancement and the energy  region of the results in this study are both smaller than those in Ref.~\cite{ZhenZhang2017} because  the effects of effective masses on $f^*$  were considered in this study but not in Ref.~\cite{ZhenZhang2017}. This suggests that a further experimental study of the pion production in heavy ion collisions will be useful for elucidating the in-medium $N\pi\to \Delta$.

\begin{figure}[htbp]
\begin{center}
\includegraphics[scale=0.3]{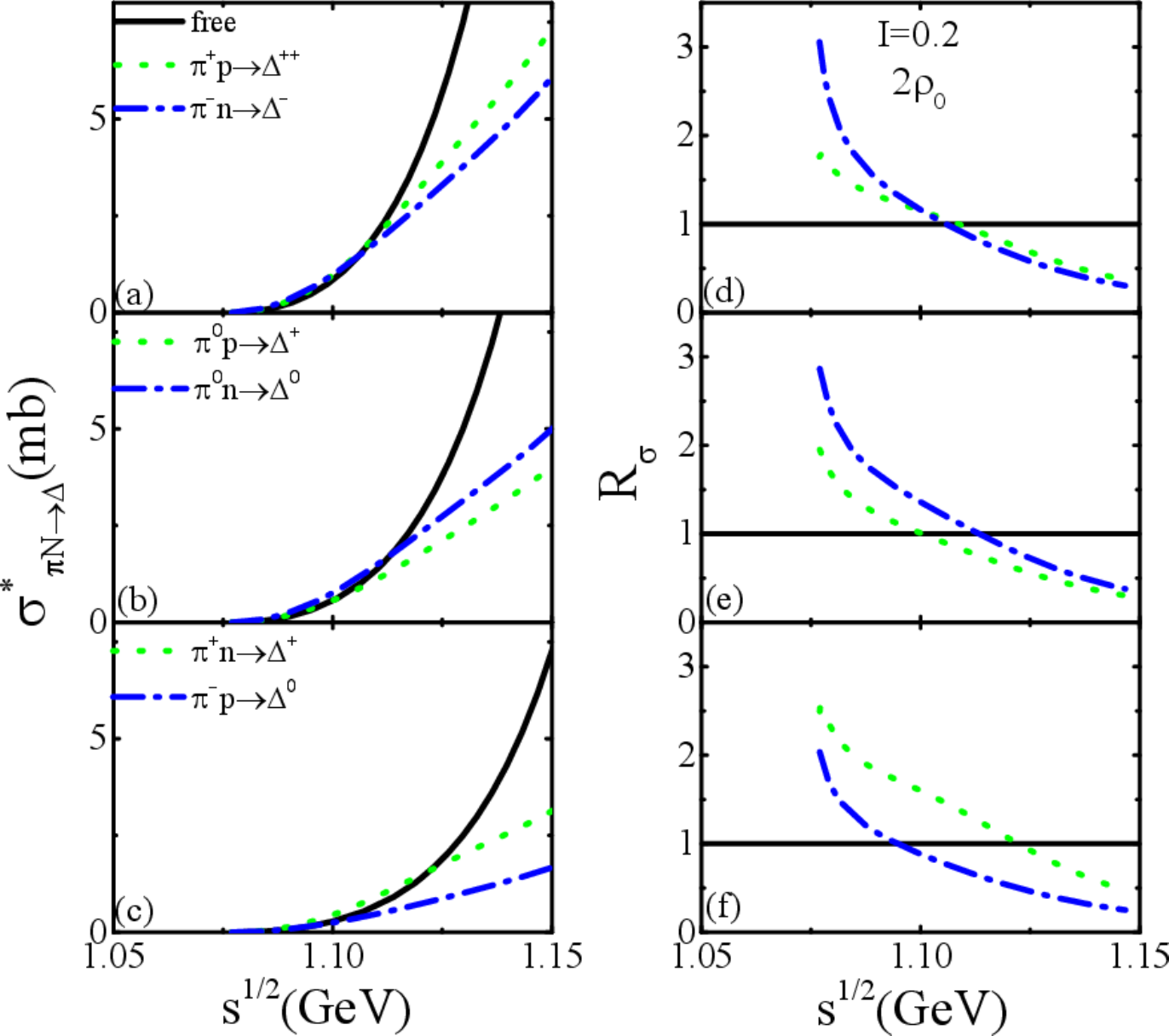}
\caption{(Color online) Left panels: the in-medium cross sections of $\pi N\to\Delta$; right panels: in-medium correction factor $R_\sigma$ at $2\rho_0$ in isospin asymmetric nuclear matter with I=0.2; $\pi^{+} p\to \Delta^{++}$ and $\pi^{-} n\to \Delta^{-}$ (upper panels), $\pi^{0} p\to \Delta^{+}$ and $\pi^{0} n\to \Delta^{0}$ (middle panels), and $\pi^{+} n\to \Delta^{+}$ and $\pi^{-} p\to \Delta^{0}$ (bottom panels).}\label{fig3v7}
\end{center}
\end{figure}

Given that the nuclear medium correction on the decay width of $\Delta \to \pi N$ is weak in the approximation adopted in this study, we next focus on the in-medium cross sections of $N\pi\to \Delta$ and their correction factor in isospin asymmetric nuclear matter. In Fig.~\ref{fig3v7}, we present the in-medium cross sections and $R_\sigma=\sigma^*/\sigma^{\mathrm{free}}$ for $\pi^{+} p\to \Delta^{++}$, $\pi^{-} n\to \Delta^{-}$, $\pi^{0} p\to \Delta^{+}$, $\pi^{0} n\to \Delta^{0}$, $\pi^{+} n\to \Delta^{+}$ and $\pi^{-} p\to \Delta^{0}$ channels  respectively  at  $2\rho_0$ in asymmetric nuclear matter with $I=0.2$. If we do not consider the  splitting of effective masses, the ratios between the cross sections of different channels are $\sigma_{\pi^{+} p\to \Delta^{++}}$ ($\sigma_{\pi^{-} n\to \Delta^{-}}$): $\sigma_{\pi^{0} p\to \Delta^{+}}$ ($\sigma_{\pi^{0} n\to \Delta^{0}}$) :$\sigma_{\pi^{+} n\to \Delta^{+}}$ ($\sigma_{\pi^{-} p\to \Delta^{0}}$)$=3:2:1$, and the medium correction factors, i.e. $R=\sigma^*/\sigma^{\mathrm{free}}$, are the same for different channels. With the nucleon and $\Delta$ effective masses splitting as well as pion energies in asymmetric nuclear matter to be considered, the in-medium correction factors $R$ on the cross sections of $\pi N\to \Delta$ and $\Delta \to \pi N $ are different for different channels as $\omega_{\pi^+}>\omega_{\pi^-}$ at higher energies, and $m^{*}_{\Delta^{++}}>m^{*}_{\Delta^{+}}>m^{*}_{\Delta^{0}}>m^{*}_{\Delta^{-}}$, and $m^{*}_{p}>m^{*}_{n}$. The cross sections of  different channels for $N\pi\to\Delta$ cross sections in asymmetric nuclear matter are  shown in the left panels of Fig.~\ref{fig3v7}. It also can be more clearly observed in the right panels of  Fig.~\ref{fig3v7}, i.e., $R_{\pi^{-} n\to \Delta^{-}}>R_{\pi^{+} p\to \Delta^{++}}$ at $s^{1/2}<1.11$ GeV while $R_{\pi^{-} n\to \Delta^{-}}<R_{\pi^{+} p\to \Delta^{++}}$ at $s^{1/2}>1.11$ GeV, $R_{\pi^{0} n\to \Delta^{0}}>R_{\pi^{0} p\to \Delta^{+}}$, and $R_{\pi^{+} n\to \Delta^{+}}>R_{\pi^{-} p\to \Delta^{0}}$ at $s^{1/2}<1.15$ GeV. 

The results are similar to those of the study from Li el at in Ref.~\cite{QingfengLi2017v2}, but the magnitude of the in-medium cross section and the splitting among the different channels are more evident than in Ref.~\cite{QingfengLi2017v2}, where the effect of $\omega$ in asymmetric nuclear matter was ignored. Based on above discussions on the in-medium cross section of $N\pi\to\Delta$, one can expect that if $\sigma^*_{N\pi\to\Delta}$ is included in transport model simulations, the production of pion may be modified, with the beam energy decreasing from 0.4 A GeV to 0.3 A GeV.

\section{Summary and outlook}
\label{summary}
In summary, we investigated the pion dispersion relation, in-medium $N\pi\to \Delta$ cross section and $\Delta \to N \pi$ decay width near the threshold energy of pion production in isospin asymmetric nuclear matter by using the same relativistic interaction within the framework of the one-boson-exchange model. With the consideration of threshold effects (or energy conservation in isospin asymmetric nuclear medium) and in-medium pion energy effects, $f^*$ is enhanced   at $s^{1/2}<1.11$ GeV and reduced at $s^{1/2}>1.11$ GeV. This results in an enhancement of in-medium $N\pi\to\Delta$ cross sections near $s^{1/2}<1.11$ GeV and then suppression at $s^{1/2}>1.11$ GeV, similar to the conclusion in Ref.~\cite{ZhenZhang2017}. Concerning the in-medium decay width of $\Delta\to N\pi$, it is reduced at $s^{1/2}<1.15$ GeV.

By including the pion energy  $\omega$ and effective mass splitting in asymmetric nuclear matter for the calculation of $N\pi\to \Delta$, our results show that the in-medium correction factors on the cross sections of $\pi N\to \Delta$ are different for different channels, e.g., $R_{\pi^{+}p\to\Delta^{++}}<R_{\pi^{-}n\to\Delta^{-}}$. As a result of the  medium correction and isospin splitting of $\sigma^*_{N\pi\to\Delta}$ in asymmetric nuclear matter, a smaller pion multiplicity and $\pi^-/\pi^+$ ratios could be predicted with respect to the calculations utility of $\sigma^{\mathrm{free}}_{N\pi\to\Delta}$  near the threshold energy if the other parameters in the transport model remain unchanged.

However, it should be kept in mind that the simulation of heavy ion collision is much more complicated. Our results suggest that a systematic study of the pion production mechanism  near the threshold energy of pion production by using multi-observables, i.e.,  pion's multiplicity, energy spectral and flow, is needed. With increase in the beam energy, there are more $\pi N\to \Delta$ and $N\Delta\to NN$ processes taking place. Thus, the width of $\Delta$ in the $\Delta$ propagator should be considered in the calculations of pion self energy, in-medium cross section of $N\pi\to\Delta$ and decay width of $\Delta$. Beam energy scanning, for example, from subthreshold energy to 1.5A GeV, and system size dependence, from smaller systems to heavier systems, could help us elucidate the medium effects on the cross sections of $N\pi\to \Delta$.


\acknowledgments
This work has been supported by National Key R\&D Program of China under Grant No. 2018 YFA0404404, and National Natural Science Foundation of China under Grants No. 11875323, No. 11875125, No. 11475262, No. 11961141003, No. 11790323, 11790324, No. 11790325, and the Continuous Basic Scientific Research Project (No. WDJC-2019-13, No BJ20002501).\\

\newpage

\begin{appendices}

\section{Appendix A}
\label{AppendixA}
\small{\begin{table}[htbp]
\begin{center}
\caption{ Isospin factors $I_{NN}$.}\label{tabt1}
\begin{tabular}{c c}
  \hline
  \hline
   $NN\pi$ & $I_{NN}$ \\
    \hline
   $pp\pi^0$ & 1\\

   $nn\pi^0$ & -1\\

  $pn\pi^+$ &-$\sqrt{2}$\\

   $np\pi^-$ & $\sqrt{2}$\\
          \hline
   \hline	 		 	 	 			 	 	
\end{tabular}
\end{center}
\end{table}}

\small{\begin{table}[htbp]
\begin{center}
\caption{Isospin factors $I_{N\Delta}$. }\label{tabd}
\begin{tabular}{c c }
  \hline
  \hline
  Channel & $I_{N\Delta}$\\
  \hline
  $\Delta^{++}\to \pi^{+}p$ & $1$ \\
 $\Delta^{+}\to \pi^{+} n$ & $\sqrt{\frac{1}{3}}$ \\
 $\Delta^{+}\to \pi^{0} p$ & $\sqrt{\frac{2}{3}}$ \\
  $\Delta^{0}\to \pi^{0} n$ & $\sqrt{\frac{2}{3}}$\\
  $\Delta^{0}\to \pi^{-} p$ & $\sqrt{\frac{1}{3}}$ \\
  $\Delta^{-}\to \pi^{-}n$ & $1$ \\
  \hline
  \hline
\end{tabular}
\end{center}
\end{table}}

\section{Appendix B}
\label{AppendixB}
Here we  remove the contributions from virtual
particle-particle excitations in Ref. \cite{Mao1999prc}, which is consistent
with the mean field approximation. According to the on-shell pion dispersion relation in Eq.~(\ref{eq:pid}), $\Pi(k)$ means the real part of pion self-energy , ($\mathrm{Re}\Pi(k)$). For convenience, "Re" ahead of  $\Pi(k)$ is ignored  in the following discussion.

\begin{figure}[htbp]
\begin{center}
\includegraphics[width=6cm]{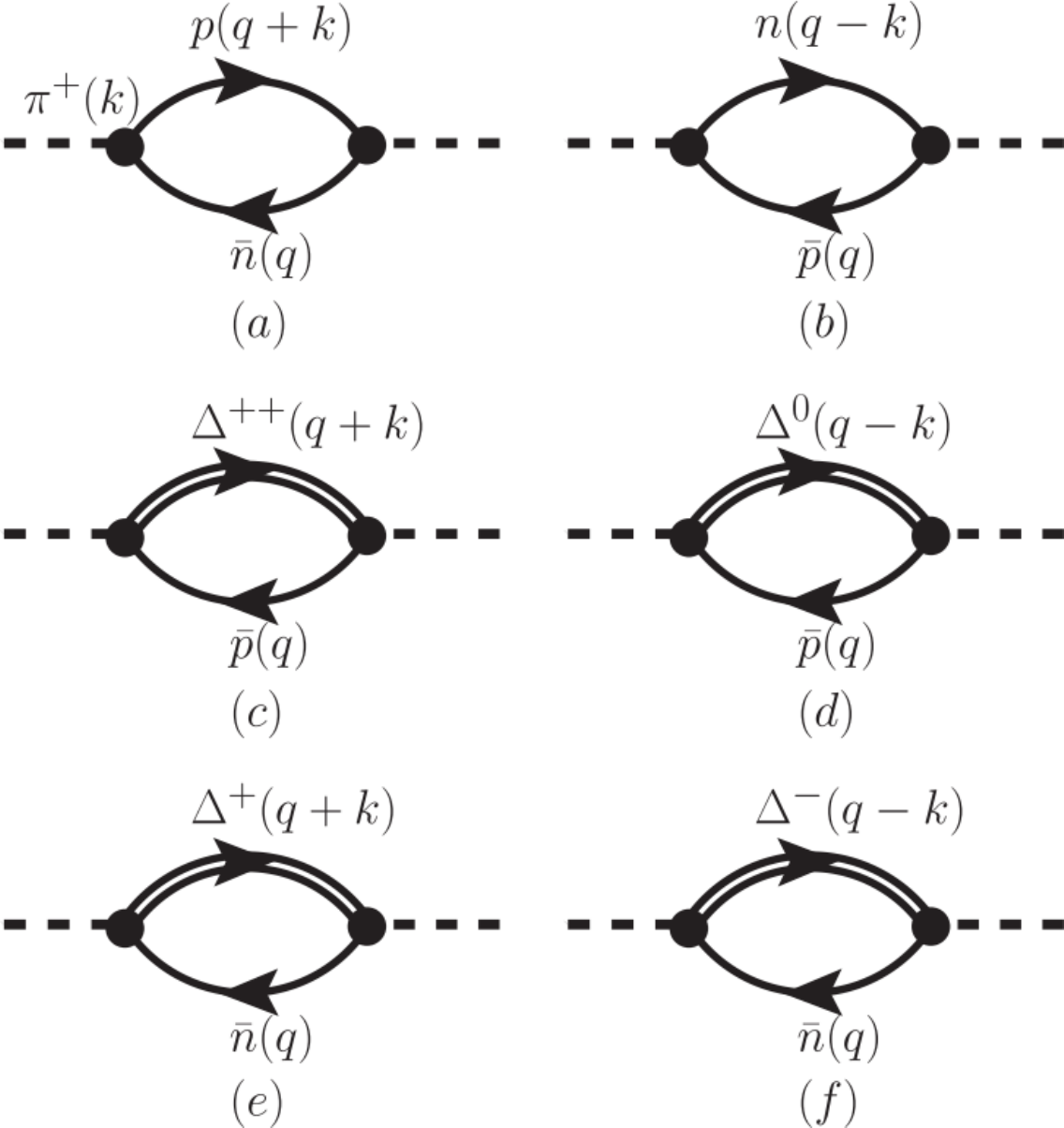}
\caption{Self-energy of $\pi^{+}(\omega, \mathbf{k})$, (a) and (b) constitute the particle-hole part, whereas (c), (d), (e), and (f) constitute the $\Delta$-hole part.}\label{figpip1ap}
\end{center}
\end{figure}
The particle-hole part of the $\pi^{+}$ self-energy can be written as follows:
\begin{equation}
\Pi_{N}(\pi^{+})=\Pi_{a}(\pi^{+})+\Pi_{b}(\pi^{+})
\end{equation}
where $\Pi_{a}(\pi^{+})$ is
\begin{eqnarray}
&&\Pi_{a}(\pi^{+})\nonumber\\
&&=-i(\frac{-\sqrt{2}g_{\pi NN}}{m_{\pi}})^2 \int\frac{d^4 q}{(2\pi)^4} \mathrm{Tr} \big [ k\!\!\!\!/\gamma_{5}
\frac{q\!\!\!/+m_{n}}{2E_{n}(q)} k\!\!\!\!/\gamma_{5} \nonumber\\
&& \times\frac{q\!\!\!\!/+k\!\!\!\!/+m_{p}}{(q_0+k_0)^2-E^{2}_{p}(q+k)}
 i2\pi\theta(q_{F, n}-|\mathbf{q}|)\delta(q^0-E_{n}(q))]\nonumber\\
&&=(\frac{g_{\pi NN}}{m_{\pi}})^2 \int\frac{d^4 q}{(2\pi)^3} \theta(q_{F, n}-|\mathbf{q}|)\delta(q^0-E_{n}(q))\nonumber\\
&&\times \frac{-4m_{n}m_{p}k^2-4q^2k^2+8(qk)^2+4k^2(q k)}{E_{n}(q)((q_0+k_0)^2-E^{2}_{p}(q+k))}\nonumber\\
&&=(\frac{g_{\pi NN}}{m_{\pi}})^2 \int\frac{d^3 \mathbf{q}}{(2\pi)^3} \frac{\theta(q_{F, n}-|\mathbf{q}|)}{E_{n}(q)[(E_{n}(q)+\omega)^2-E^{2}_{p}(q+k)]}\nonumber\\
&&\times [-4m_{n}m_{p}k^2-4m^{2}_{n}k^2\nonumber\\
&&+4(E_{n}(q)\omega-\mathbf{q}\cdot \mathbf{k})(2E_{n}(q)\omega-2\mathbf{q}\cdot \mathbf{k}+k^2) ].\nonumber\\
\end{eqnarray}
Here, $k_0=\omega$, $E_{n}(q)=\sqrt{m^2_n+\mathbf{q}^2}$ and  $k^2=k^2_0-\mathbf{k}^2=\omega^2-\mathbf{k}^2$.  In addition, $n(|\mathbf{q}|)=\theta(q_F-|\mathbf{q}|)$
denotes the  occupation number in zero temperature  nuclear matter in the Fermi momentum $q_F$. The isospin factor $I_{N\Delta}=-\sqrt{2}$ is listed in Table \ref{tabt1}.  Finally, $\Pi_{b}(\pi^{+})$ can also be calculated in the same way:
\begin{eqnarray}
&&\Pi_{b}(\pi^{+})\nonumber\\
&&=-i(\frac{\sqrt{2}g_{\pi NN}}{m_{\pi}})^2 \int\frac{d^4 q}{(2\pi)^4} \mathrm{Tr} \big [ k\!\!\!\!/\gamma_{5} \frac{q\!\!\!\!/+m_{p}}{2E_{p}(q)} \nonumber\\
&&\times k\!\!\!\!/\gamma_{5}\frac{q\!\!\!\!/-k\!\!\!\!/+m_{n}}{(q_0-k_0)^2-E^{2}_{n}(q-k)}\nonumber\\
&& \times i2\pi\theta(q_{F, p}-|\mathbf{q}|)\delta(q^0-E_{p}(q))]\\
&&=(\frac{g_{\pi NN}}{m_{\pi}})^2 \int\frac{d^3 \mathbf{q}}{(2\pi)^3} \frac{\theta(q_{F, p}-|\mathbf{q}|)}{E_{p}(q)[(E_{p}(q)-\omega)^2-E^{2}_{n}(q-k)]} \nonumber\\
&&\times[-4m_{n}m_{p}k^2-4m^{2}_{p}k^2\nonumber\\
&&+4(-E_{p}(q)\omega+\mathbf{q}\cdot \mathbf{k})(-2E_{p}(q)\omega+2\mathbf{q}\cdot \mathbf{k}+k^2) ].\nonumber\\
\end{eqnarray}
The $\Delta$-hole part of the $\pi^+$ self-energy is expressed as
\begin{equation}
\Pi_{\Delta}(\pi^{+})=\Pi_{c}(\pi^{+})+\Pi_{d}(\pi^{+})+\Pi_{e}(\pi^{+})+\Pi_{f}(\pi^{+}),
\end{equation}
where $\Pi_{c}(\pi^{+})$
\begin{eqnarray}
&&\Pi_{c}(\pi^{+})\nonumber\\
&&=-i(\frac{g_{\pi N\Delta}}{m_{\pi}})^2\int\frac{d^4 q}{(2\pi)^4} \mathrm{Tr}\big [\frac{k_{\mu}k_{\nu}D^{\mu\nu}(q+k)(q\!\!\!\!/+k\!\!\!\!/+m_{0,\Delta})}{(q_0+k_0)^2-E^{2}_{\Delta}(q+k)}\nonumber\\
&&\times \frac{q\!\!\!\!/+m_{n}}{2E_{p}(q)}\theta(q_{F, p}-|\mathbf{q}|)(i2\pi\delta(q_0-E_{p}(q)))\big]\nonumber\\
&&=(\frac{g_{\pi N\Delta}}{m_{\pi}})^2\int\frac{d^3 \mathbf{q}}{(2\pi)^3} \frac{\theta(q_{F, p}-|\mathbf{q}|)}{2E_{p}(q)((E_{p}(q)+k_0)^2-E^{2}_{\Delta}(q+k))}\nonumber\\
&&\times 4[\frac{2m_{p}(qk)^2}{3m_{0,\Delta^{++}}}+\frac{4m_{p}(qk)k^2}{3m_{0,\Delta^{++}}}+\frac{2m_{p}k^4}{3m_{0,\Delta^{++}}}-\frac{2m_{p}m_{0,\Delta^{++}}k^2}{3}\nonumber\\
&&+\frac{2q^2 k^4}{3m^{2}_{0,\Delta^{++}}}+\frac{2(q k)^3}{3m^{2}_{0,\Delta^{++}}} +\frac{2q^2 (q k)^2}{3m^{2}_{0,\Delta^{++}}}+\frac{4k^2 (q k)^2}{3m^{2}_{0,\Delta^{++}}}\nonumber\\
&&+\frac{2k^4 (q k)}{3m^{2}_{0,\Delta^{++}}}+\frac{4q^2k^2 (q k)}{3m^{2}_{0,\Delta^{++}}}-\frac{2q^2k^2 }{3}-\frac{2k^2(pk) }{3}]\nonumber\\
&&=\frac{2}{3}(\frac{g_{\pi N\Delta}}{m_{\pi}})^2\int\frac{d^3 \mathbf{q}}{(2\pi)^3} \frac{\theta(q_{F, p}-|\mathbf{q}|)}{E_{p}(q)}\nonumber\\
&&\times\large [\frac{(qk)^2-m^2_{p}k^2}{m^{2}_{0,\Delta^{++}}}+k^2\frac{2m_{p}}{m_{0,\Delta^{++}}}(1+\frac{m_{p}}{m_{0,\Delta^{++}}})\nonumber\\
&&+\frac{(qk)^2-m^{2}_{p}k^2}{m^{2}_{0,\Delta^{++}}}\frac{(m_{p}+m_{0,\Delta^{++}})^2-k^2}{2qk+k^2-(m^{2}_{0,\Delta^{++}}-m^{2}_{p})}\large],\nonumber\\
\end{eqnarray}
where $qk=E_{p}(q)\omega-\mathbf{q}\cdot \mathbf{k}$.
$\Pi_{d}(\pi^{+})$, $\Pi_{e}(\pi^{+})$, and $\Pi_{f}(\pi^{+})$ can be obtained in the same way:
\begin{eqnarray}
&&\Pi_{d}(\pi^{+})\nonumber\\
&&=\frac{2}{9}(\frac{g_{\pi N\Delta}}{m_{\pi}})^2\int\frac{d^3 \mathbf{q}}{(2\pi)^3} \frac{\theta(q_{F, p}-|\mathbf{q}|)}{E_{p}(q)}\nonumber\\
&&\times\large [\frac{(qk)^2-m^2_{p}k^2}{m^{2}_{0,\Delta^{0}}}+k^2\frac{2m_{p}}{m_{0,\Delta^{0}}}(1+\frac{m_{p}}{m_{0,\Delta^{0}}})\nonumber\\
&&+\frac{(qk)^2-m^{2}_{p}k^2}{m^{2}_{0,\Delta^{0}}}\frac{(m_{p}+m_{0,\Delta^{0}})^2-k^2}{-2qk+k^2-(m^{2}_{0,\Delta^{0}}-m^{2}_{p})}\large],\nonumber\\
\end{eqnarray}

\begin{eqnarray}
&&\Pi_{e}(\pi^{+})\nonumber\\
&&=\frac{2}{9}(\frac{g_{\pi N\Delta}}{m_{\pi}})^2\int\frac{d^3 \mathbf{q}}{(2\pi)^3} \frac{\theta(q_{F, n}-|\mathbf{q}|)}{E_{n}(q)}\nonumber\\
&&\times\large [\frac{(qk)^2-m^2_{n}k^2}{m^{2}_{0,\Delta^{+}}}+k^2\frac{2m_{n}}{m_{0,\Delta^{+}}}(1+\frac{m_{n}}{m_{0,\Delta^{+}}})\nonumber\\
&&+\frac{(qk)^2-m^{2}_{n}k^2}{m^{2}_{0,\Delta^{+}}}\frac{(m_{n}+m_{0,\Delta^{+}})^2-k^2}{2qk+k^2-(m^{2}_{0,\Delta^{+}}-m^{2}_{n})}\large],\nonumber\\
\end{eqnarray}

\begin{eqnarray}
&&\Pi_{f}(\pi^{+})\nonumber\\
&&=\frac{2}{3}(\frac{g_{\pi N\Delta}}{m_{\pi}})^2\int\frac{d^3 \mathbf{q}}{(2\pi)^3} \frac{\theta(q_{F, n}-|\mathbf{q}|)}{E_{n}(q)}\nonumber\\
&&\times\large [\frac{(qk)^2-m^2_{n}k^2}{m^{2}_{0,\Delta^{-}}}+k^2\frac{2m_{n}}{m_{0,\Delta^{-}}}(1+\frac{m_{n}}{m_{0,\Delta^{-}}})\nonumber\\
&&+\frac{(qk)^2-m^{2}_{n}k^2}{m^{2}_{0,\Delta^{-}}}\frac{(m_{n}+m_{0,\Delta^{-}})^2-k^2}{-2qk+k^2-(m^{2}_{0,\Delta^{-}}-m^{2}_{n})}\large].\nonumber\\
\end{eqnarray}

\begin{figure}[htbp]
\begin{center}
\includegraphics[width=6cm]{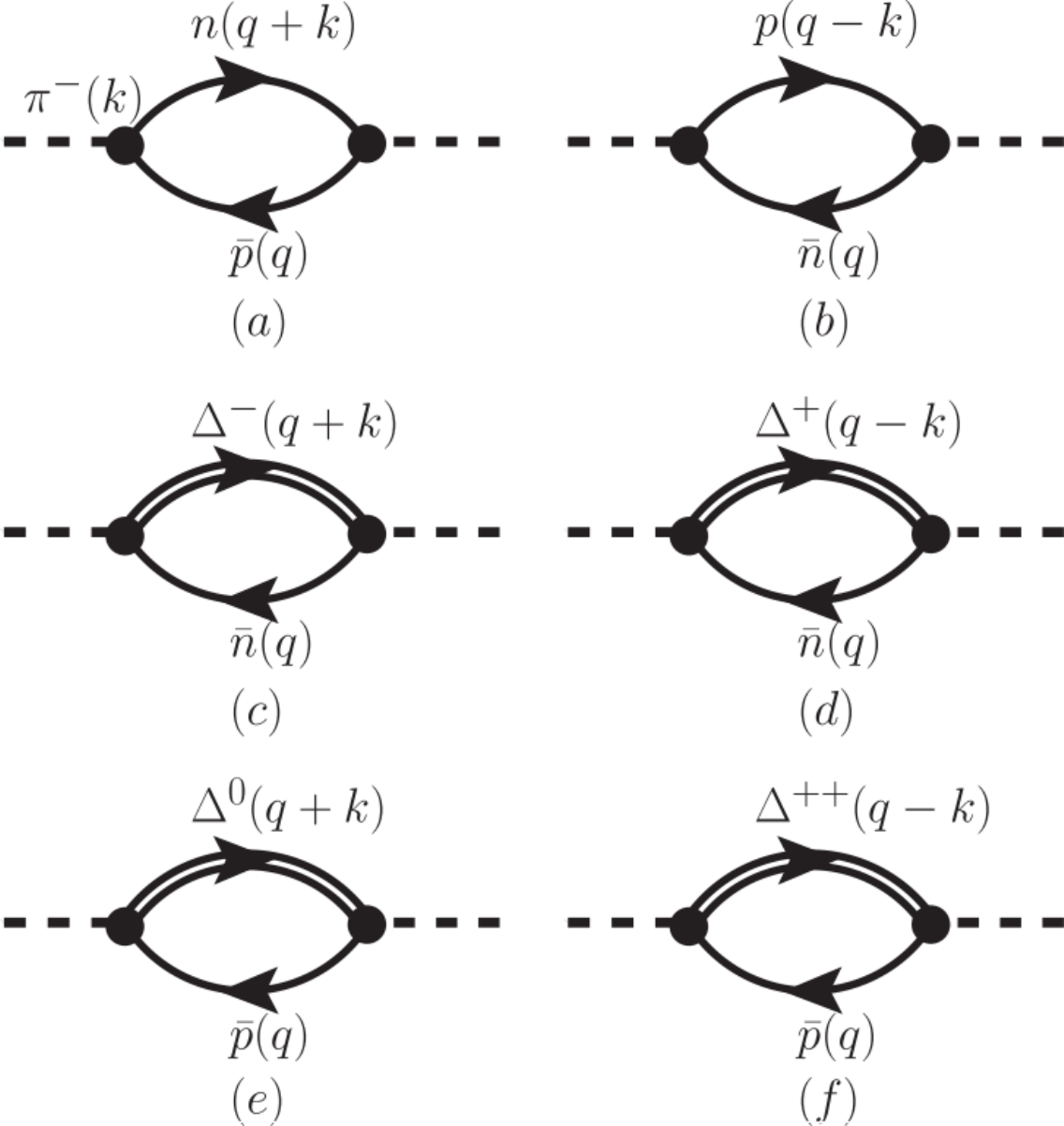}
\caption{$\pi^{-}(\omega, \mathbf{k})$ self-energy}\label{figpim1ap}
\end{center}
\end{figure}

The particle-hole part of the $\pi^{-}$ self-energy can be written as follows:
\begin{equation}
\Pi_{N}(\pi^{-})=\Pi_{a}(\pi^{-})+\Pi_{b}(\pi^{-})
\end{equation}
Here, $\Pi_{a}(\pi^{-})$ and $\Pi_{b}(\pi^{-})$ are expressed as follows:
\begin{eqnarray}
&&\Pi_{a}(\pi^{-})\nonumber\\
&&=(\frac{g_{\pi NN}}{m_{\pi}})^2 \int\frac{d^3 \mathbf{q}}{(2\pi)^3} \frac{\theta(q_{F, p}-|\mathbf{q}|)}{E_{p}(q)[(E_{p}(q)+\omega)^2-E^{2}_{n}(q+k)]}\nonumber\\
&&\times [-4m_{n}m_{p}k^2-4m^{2}_{p}k^2\nonumber\\
&&+4(E_{p}(q)\omega-\mathbf{q}\cdot \mathbf{k})(2E_{p}(q)\omega-2\mathbf{q}\cdot \mathbf{k}+k^2) ].
\end{eqnarray}

\begin{eqnarray}
&&\Pi_{b}(\pi^{-})\nonumber\\
&&=(\frac{g_{\pi NN}}{m_{\pi}})^2 \int\frac{d^3 \mathbf{q}}{(2\pi)^3} \frac{\theta(q_{F, n}-|\mathbf{q}|)}{E_{n}(q)[(E_{n}(q)-\omega)^2-E^{2}_{p}(q-k)]}\nonumber\\
&&\times [-4m_{n}m_{p}k^2-4m^{2}_{n}k^2\nonumber\\
&&+4(-E_{n}(q)\omega+\mathbf{q}\cdot \mathbf{k})(-2E_{n}(q)\omega+2\mathbf{q}\cdot \mathbf{k}+k^2) ].
\end{eqnarray}

The $\Delta$-hole part of the $\pi^-$ self-energy is
\begin{equation}
\Pi_{\Delta}(\pi^{-})=\Pi_{c}(\pi^{-})+\Pi_{d}(\pi^{-})+\Pi_{e}(\pi^{-})+\Pi_{f}(\pi^{-})
\end{equation}
Here, $\Pi_{c}(\pi^{-})$, $\Pi_{d}(\pi^{-})$, $\Pi_{e}(\pi^{-})$ and $\Pi_{f}(\pi^{-})$  can be calculated as follows:
\begin{eqnarray}
&&\Pi_{c}(\pi^{-})\nonumber\\
&&=\frac{2}{3}(\frac{g_{\pi N\Delta}}{m_{\pi}})^2\int\frac{d^3 \mathbf{q}}{(2\pi)^3} \frac{\theta(q_{F, n}-|\mathbf{q}|)}{E_{n}(q)}\nonumber\\
&&\times\large [\frac{(qk)^2-m^2_{n}k^2}{m^{2}_{0,\Delta^{-}}}+k^2\frac{2m_{n}}{m_{0,\Delta^{-}}}(1+\frac{m_{n}}{m_{0,\Delta^{-}}})\nonumber\\
&&+\frac{(qk)^2-m^{2}_{n}k^2}{m^{2}_{0,\Delta^{-}}}\frac{(m_{n}+m_{0,\Delta^{-}})^2-k^2}{2qk+k^2-(m^{2}_{0,\Delta^{-}}-m^{2}_{n})}\large],
\end{eqnarray}

\begin{eqnarray}
&&\Pi_{d}(\pi^{-})\nonumber\\
&&=\frac{2}{9}(\frac{g_{\pi N\Delta}}{m_{\pi}})^2\int\frac{d^3 \mathbf{q}}{(2\pi)^3} \frac{\theta(q_{F, n}-|\mathbf{q}|)}{E_{n}(q)}\nonumber\\
&&\times\large [\frac{(qk)^2-m^2_{n}k^2}{m^{2}_{0,\Delta^{+}}}+k^2\frac{2m_{n}}{m_{0,\Delta^{+}}}(1+\frac{m_{n}}{m_{0,\Delta^{+}}})\nonumber\\
&&+\frac{(qk)^2-m^{2}_{n}k^2}{m^{2}_{0,\Delta^{+}}}\frac{(m_{n}+m_{0,\Delta^{+}})^2-k^2}{-2qk+k^2-(m^{2}_{0,\Delta^{+}}-m^{2}_{n})}\large],\nonumber\\
\end{eqnarray}

\begin{eqnarray}
&&\Pi_{e}(\pi^{-})\nonumber\\
&&=\frac{2}{9}(\frac{g_{\pi N\Delta}}{m_{\pi}})^2\int\frac{d^3 \mathbf{q}}{(2\pi)^3} \frac{\theta(q_{F, p}-|\mathbf{q}|)}{E_{p}(q)}\nonumber\\
&&\times\large [\frac{(qk)^2-m^2_{p}k^2}{m^{2}_{0,\Delta^{0}}}+k^2\frac{2m_{p}}{m_{0,\Delta^{0}}}(1+\frac{m_{p}}{m_{0,\Delta^{0}}})\nonumber\\
&&+\frac{(qk)^2-m^{2}_{p}k^2}{m^{2}_{0,\Delta^{0}}}\frac{(m_{p}+m_{0,\Delta^{0}})^2-k^2}{2qk+k^2-(m^{2}_{0,\Delta^{0}}-m^{2}_{p})}\large],\nonumber\\
\end{eqnarray}

\begin{eqnarray}
&&\Pi_{f}(\pi^{-})\nonumber\\
&&=\frac{2}{3}(\frac{g_{\pi N\Delta}}{m_{\pi}})^2\int\frac{d^3 \mathbf{q}}{(2\pi)^3} \frac{\theta(q_{F, p}-|\mathbf{q}|)}{E_{p}(q)}\nonumber\\
&&\times\large [\frac{(qk)^2-m^2_{p}k^2}{m^{2}_{0,\Delta^{++}}}+k^2\frac{2m_{p}}{m_{0,\Delta^{+}}}(1+\frac{m_{p}}{m_{0,\Delta^{++}}})\nonumber\\
&&+\frac{(qk)^2-m^{2}_{p}k^2}{m^{2}_{0,\Delta^{++}}}\frac{(m_{p}+m_{0,\Delta^{++}})^2-k^2}{-2qk+k^2-(m^{2}_{0,\Delta^{++}}-m^{2}_{p})}\large].\nonumber\\
\end{eqnarray}

\begin{figure}[htbp]
\begin{center}
\includegraphics[width=6cm]{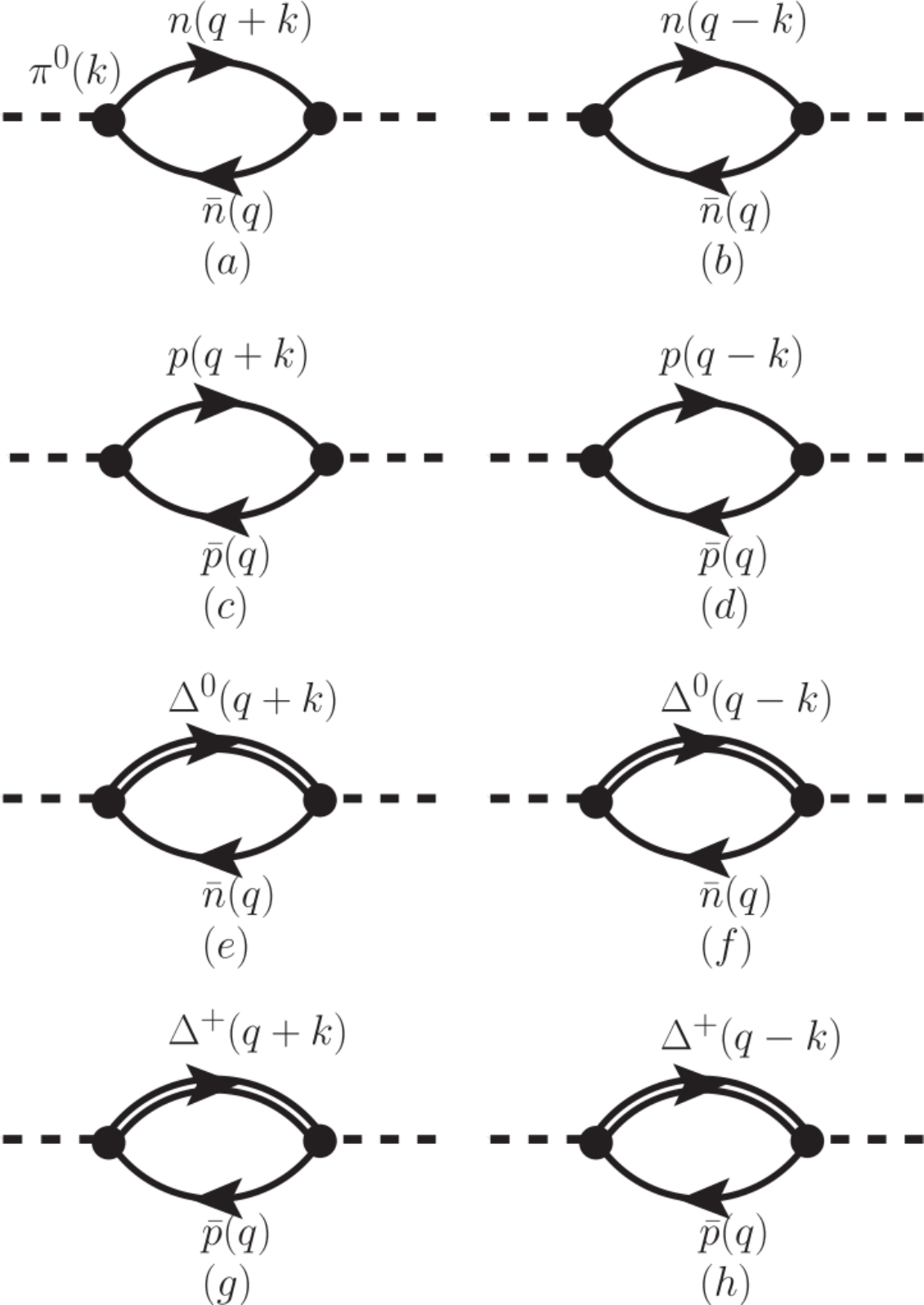}
\caption{$\pi^{0}(\omega, \mathbf{k})$ self-energy}\label{figpi01ap}
\end{center}
\end{figure}

The particle-hole part of the $\pi^{0}$ self-energy can be written as follows:
\begin{equation}
\Pi_{N}(\pi^{0})=\Pi_{a}(\pi^{0})+\Pi_{b}(\pi^{0})+\Pi_{c}(\pi^{0})+\Pi_{d}(\pi^{0}).
\end{equation}
Here, $\Pi_{a}(\pi^{0})$, $\Pi_{b}(\pi^{0})$, $\Pi_{c}(\pi^{0})$, and $\Pi_{d}(\pi^{0})$ are expressed as follows:

\begin{eqnarray}
&&\Pi_{a}(\pi^0) \nonumber\\
&&=(\frac{g_{\pi NN}}{m_{\pi}})^2 \int\frac{d^3 \mathbf{q}}{(2\pi)^3}\theta(q_{F, n}-|\bm{q}|) \nonumber\\
&&\times [\frac{-4m^{2}_{n}k^2}{E_{n}(q)(2E_{n}(q)\omega-2\mathbf{q}\cdot \mathbf{k}+\omega^2-\mathbf{k}^2)} +2\omega ] ,\nonumber\\
\end{eqnarray}

\begin{eqnarray}
&&\Pi_{b}(\pi^0) \nonumber\\
&&=(\frac{g_{\pi NN}}{m_{\pi}})^2 \int\frac{d^3 \mathbf{q}}{(2\pi)^3}\theta(q_{F, n}-|\bm{q}|) \nonumber\\
&&\times [\frac{-4m^{2}_{n}k^2}{E_{n}(q)(-2E_{n}(q)\omega+2\mathbf{q}\cdot \mathbf{k}+\omega^2-\mathbf{k}^2)} -2\omega ] ,\nonumber\\
\end{eqnarray}

\begin{eqnarray}
&&\Pi_{c}(\pi^0) \nonumber\\
&&=(\frac{g_{\pi NN}}{m_{\pi}})^2 \int\frac{d^3 \mathbf{q}}{(2\pi)^3}\theta(q_{F, p}-|\bm{q}|) \nonumber\\
&&\times [\frac{-4m^{2}_{p}k^2}{E_{p}(q)(2E_{p}(q)\omega-2\mathbf{q}\cdot \mathbf{k}+\omega^2-\mathbf{k}^2)} +2\omega ] ,\nonumber\\
\end{eqnarray}

\begin{eqnarray}
&&\Pi_{d}(\pi^0) \nonumber\\
&&=(\frac{g_{\pi NN}}{m_{\pi}})^2 \int\frac{d^3 \mathbf{q}}{(2\pi)^3}\theta(q_{F, p}-|\bm{q}|) \nonumber\\
&&\times [\frac{-4m^{2}_{p}k^2}{E_{p}(q)(-2E_{p}(q)\omega+2\mathbf{q}\cdot \mathbf{k}+\omega^2-\mathbf{k}^2)} -2\omega ].\nonumber\\
\end{eqnarray}

The $\Delta$-hole part of the $\pi^-$ self-energy is
\begin{equation}
\Pi_{\Delta}(\pi^{0})=\Pi_{e}(\pi^{0})+\Pi_{f}(\pi^{0})+\Pi_{g}(\pi^{0})+\Pi_{h}(\pi^{0}).
\end{equation}
Here, $\Pi_{e}(\pi^{0})$, $\Pi_{f}(\pi^{0})$, $\Pi_{g}(\pi^{0})$, and $\Pi_{h}(\pi^{0})$  can be calculated as follows:
\begin{eqnarray}
&&\Pi_{e}(\pi^{0})\nonumber\\
&&=\frac{4}{9}(\frac{g_{\pi N\Delta}}{m_{\pi}})^2\int\frac{d^3 \mathbf{q}}{(2\pi)^3} \frac{\theta(q_{F, n}-|\mathbf{q}|)}{E_{n}(q)}\nonumber\\
&&\times\large [\frac{(qk)^2-m^2_{n}k^2}{m^{2}_{0,\Delta^{0}}}+k^2\frac{2m_{n}}{m_{0,\Delta^{0}}}(1+\frac{m_{n}}{m_{0,\Delta^{0}}})\nonumber\\
&&+\frac{(qk)^2-m^{2}_{n}k^2}{m^{2}_{0,\Delta^{0}}}\frac{(m_{n}+m_{0,\Delta^{0}})^2-k^2}{2qk+k^2-(m^{2}_{0,\Delta^{0}}-m^{2}_{n})}\large],
\end{eqnarray}

\begin{eqnarray}
&&\Pi_{f}(\pi^{0})\nonumber\\
&&=\frac{4}{9}(\frac{g_{\pi N\Delta}}{m_{\pi}})^2\int\frac{d^3 \mathbf{q}}{(2\pi)^3} \frac{\theta(q_{F, n}-|\mathbf{q}|)}{E_{n}(q)}\nonumber\\
&&\times\large [\frac{(qk)^2-m^2_{n}k^2}{m^{2}_{0,\Delta^{0}}}+k^2\frac{2m_{n}}{m_{0,\Delta^{0}}}(1+\frac{m_{n}}{m_{0,\Delta^{0}}})\nonumber\\
&&+\frac{(qk)^2-m^{2}_{n}k^2}{m^{2}_{0,\Delta^{0}}}\frac{(m_{n}+m_{0,\Delta^{0}})^2-k^2}{-2qk+k^2-(m^{2}_{0,\Delta^{0}}-m^{2}_{n})}\large],
\end{eqnarray}

\begin{eqnarray}
&&\Pi_{g}(\pi^{0})\nonumber\\
&&=\frac{4}{9}(\frac{g_{\pi N\Delta}}{m_{\pi}})^2\int\frac{d^3 \mathbf{q}}{(2\pi)^3} \frac{\theta(q_{F, p}-|\mathbf{q}|)}{E_{p}(q)}\nonumber\\
&&\times\large [\frac{(qk)^2-m^2_{p}k^2}{m^{2}_{0,\Delta^{+}}}+k^2\frac{2m_{p}}{m_{0,\Delta^{+}}}(1+\frac{m_{p}}{m_{0,\Delta^{+}}})\nonumber\\
&&+\frac{(qk)^2-m^{2}_{p}k^2}{m^{2}_{0,\Delta^{+}}}\frac{(m_{p}+m_{0,\Delta^{+}})^2-k^2}{2qk+k^2-(m^{2}_{0,\Delta^{+}}-m^{2}_{p})}\large],
\end{eqnarray}

\begin{eqnarray}
&&\Pi_{h}(\pi^{0})\nonumber\\
&&=\frac{4}{9}(\frac{g_{\pi N\Delta}}{m_{\pi}})^2\int\frac{d^3 \mathbf{q}}{(2\pi)^3} \frac{\theta(q_{F, p}-|\mathbf{q}|)}{E_{p}(q)}\nonumber\\
&&\times\large [\frac{(qk)^2-m^2_{p}k^2}{m^{2}_{0,\Delta^{+}}}+k^2\frac{2m_{p}}{m_{0,\Delta^{+}}}(1+\frac{m_{p}}{m_{0,\Delta^{+}}})\nonumber\\
&&+\frac{(qk)^2-m^{2}_{p}k^2}{m^{2}_{0,\Delta^{+}}}\frac{(m_{p}+m_{0,\Delta^{+}})^2-k^2}{-2qk+k^2-(m^{2}_{0,\Delta^{+}}-m^{2}_{p})}\large].
\end{eqnarray}
In symmetric nuclear matter, the pion self-energy is
\begin{eqnarray}
&&\Pi_{N}\nonumber\\
&&=-8m^{2}_{N}k^2(\frac{g_{\pi NN}}{m_{\pi}})^2 \int\frac{d^3 \mathbf{q}}{(2\pi)^3}\frac{\theta(q_{F, N}-|\bm{q}|)}{E_{N}(q)} \nonumber\\
&&\times [\frac{1}{2E_{N}(q)\omega-2\mathbf{q}\cdot \mathbf{k}+\omega^2-\mathbf{k}^2} \nonumber\\
&&-\frac{1}{-2E_{N}(q)\omega+2\mathbf{q}\cdot \mathbf{k}+\omega^2-\mathbf{k}^2}  ] ,\nonumber\\
\end{eqnarray}

and
\begin{eqnarray}
&&\Pi_{\Delta}\nonumber\\
&&=\frac{8}{9}(\frac{g_{\pi N\Delta}}{m_{\pi}})^2\int\frac{d^3 \mathbf{q}}{(2\pi)^3} \frac{\theta(q_{F, N}-|\mathbf{q}|)}{E_{N}(q)}\nonumber\\
&&\times\{\frac{(qk)^2-m^2_{N}k^2}{m^{2}_{0,\Delta}}+k^2\frac{2m_{N}}{m_{0,\Delta}}(1+\frac{m_{N}}{m_{0,\Delta}})\nonumber\\
&&+\frac{(qk)^2-m^{2}_{N}k^2}{m^{2}_{0,\Delta}}[\frac{(m_{N}+m_{0,\Delta})^2-k^2}{2qk+k^2-(m^{2}_{0,\Delta}-m^{2}_{N})}\nonumber\\
&&   +\frac{(m_{N}+m_{0,\Delta})^2-k^2}{-2qk+k^2-(m^{2}_{0,\Delta}-m^{2}_{N})}  ] \}.\nonumber\\
\end{eqnarray}

\section{Appendix C}
\label{AppendixC}
The pion self-energies in Appendix A can be  expressed in terms of an analog of the
 susceptibility $\chi$ as follows:
\begin{eqnarray}
\Pi_N=k^2\chi_N\\
\Pi_\Delta=k^2\chi_\Delta.
\end{eqnarray}
Here, we introduce the effect of nonrelativistic interaction as the nuclear spin-isospin short range correlation, as in Ref.\cite{Dmitriev1985}:
\begin{eqnarray}
\label{eq:Lagsr}
W&&=(\frac{g_{\pi NN}}{m_{\pi}})^2 g^{\prime}_{NN}\bm{\sigma}_1 \cdot \bm{\sigma}_2\bm{\tau}_1\cdot\bm{\tau}_2\nonumber\\
&&+(\frac{g_{\pi N\Delta}}{m_{\pi}})^2 g^{\prime}_{\Delta\Delta}\bm{S}_1^\dagger \cdot \bm{S}_2\bm{\mathcal{T}}_1^\dagger \cdot\bm{\mathcal{T}}_2\nonumber\\
&&+\frac{g_{\pi NN}g_{\pi N\Delta}}{m_{\pi}^2} g^{\prime}_{N\Delta}\bm{S}_1^\dagger \cdot \bm{\sigma}_2\bm{\mathcal{T}}_1^\dagger \cdot\bm{\tau}_2+h.c.
\end{eqnarray}

With the short-range interaction, the pion dispersion relation can be expressed as follows:
\begin{eqnarray}
\omega^2=m_{\pi}^2+\mathbf{k}^2+\Pi=m_{\pi}^2+\mathbf{k}^2+k^2\chi
\end{eqnarray}

\begin{figure}[htbp]
\begin{center}
\includegraphics[width=7.0cm]{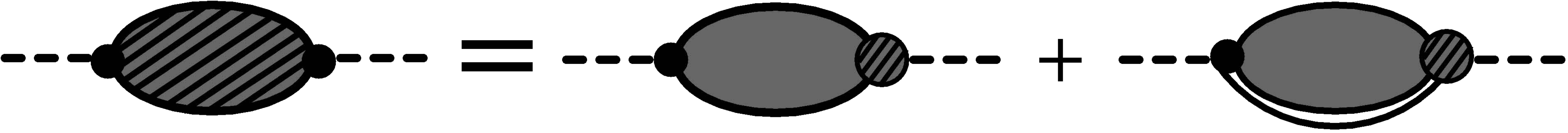}
\end{center}
\end{figure}

\begin{eqnarray}
\chi=\chi_1 +\chi_2
\end{eqnarray}

\begin{figure}[htbp]
\begin{center}
\includegraphics[width=7cm]{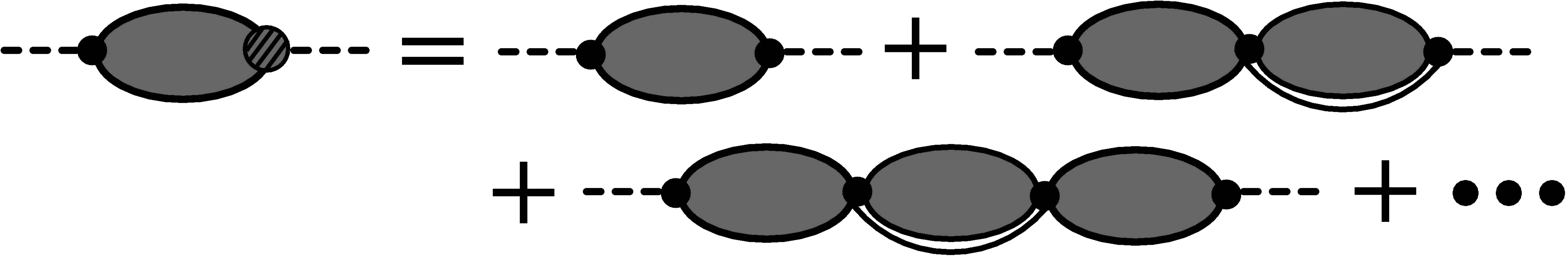}
\end{center}
\end{figure}

\begin{eqnarray}
\chi_1=\chi_{N}^{\prime}\frac{1+g^{\prime}_{N\Delta}\chi_{\Delta}^{\prime}}{1-g^{\prime}_{N\Delta}\chi_{\Delta}^{\prime}g^{\prime}_{N\Delta}\chi_{N}^{\prime}}
\end{eqnarray}

\begin{eqnarray}
\chi_2=\chi_{\Delta}^{\prime}\frac{1+g^{\prime}_{N\Delta}\chi_{N}^{\prime}}{1-g^{\prime}_{N\Delta}\chi_{N}^{\prime}g^{\prime}_{N\Delta}\chi_{\Delta}^{\prime}}.
\end{eqnarray}

\begin{figure}[htbp]
\begin{center}
\includegraphics[width=7.0cm]{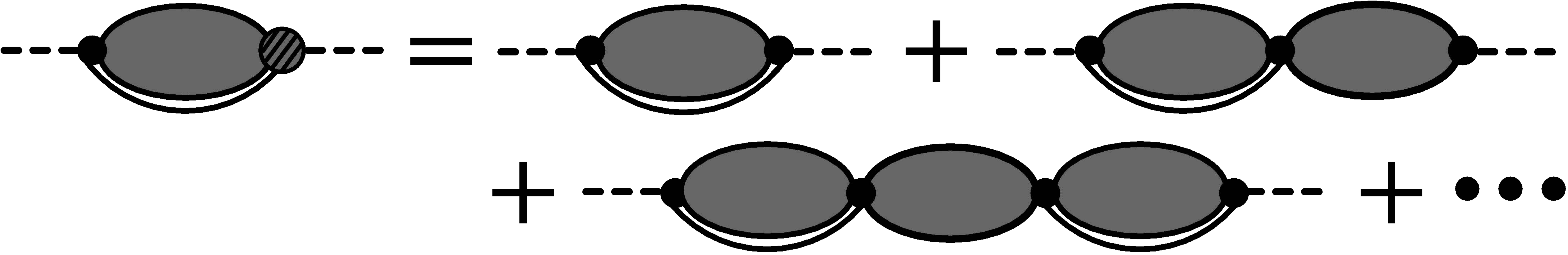}
\end{center}
\end{figure}

\begin{figure}[H]
\begin{center}
\includegraphics[width=7.0cm]{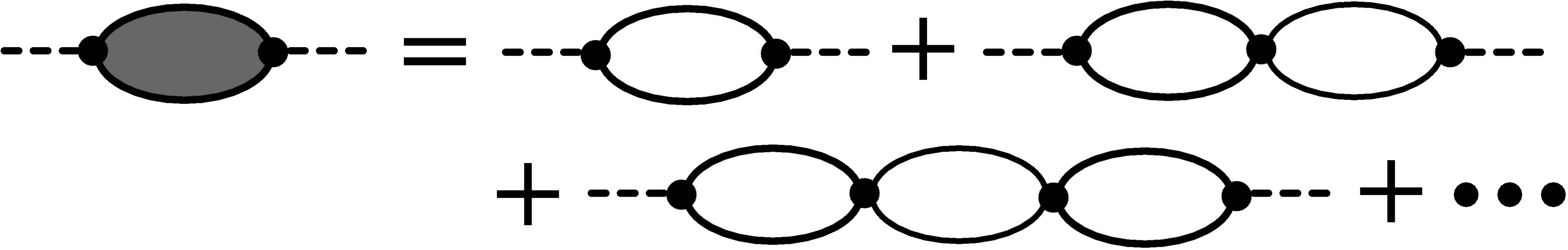}
\end{center}
\end{figure}

\begin{eqnarray}
\chi_{N}^{\prime}=\frac{\chi_{N}}{1-g^{\prime}_{NN}\chi_{N}}
\end{eqnarray}

\begin{figure}[H]
\begin{center}
\includegraphics[width=7.0cm]{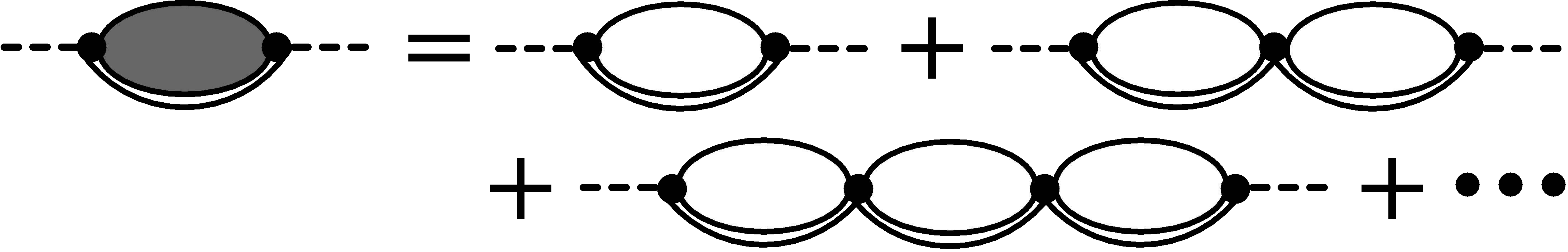}
\end{center}
\end{figure}

\begin{eqnarray}
\chi_{\Delta}^{\prime}=\frac{\chi_{\Delta}}{1-g^{\prime}_{\Delta\Delta}\chi_{\Delta}} .
\end{eqnarray}

Then, the susceptibility can be written as follows:

\begin{eqnarray}
\chi_{N}\to \chi_1=\frac{1+(g^{\prime}_{N\Delta}-g^{\prime}_{\Delta\Delta})\chi_{\Delta}}{(1-g^{\prime}_{\Delta\Delta}\chi_{\Delta})(1-g^{\prime}_{NN}\chi_{N})-g^{\prime}_{N\Delta}\chi_{\Delta}g^{\prime}_{N\Delta}\chi_{N}} \chi_{N}\nonumber\\
\chi_{\Delta}\to \chi_2=\frac{1+(g^{\prime}_{N\Delta}-g^{\prime}_{NN})\chi_{N}}{(1-g^{\prime}_{\Delta\Delta}\chi_{\Delta})(1-g^{\prime}_{NN}\chi_{N})-g^{\prime}_{N\Delta}\chi_{\Delta}g^{\prime}_{N\Delta}\chi_{N}} \chi_{\Delta}\nonumber\\
\end{eqnarray}

\section{Appendix D}
\label{AppendixD}

The pion energy $\omega$ in the nuclear matter can be written as the
parametrization form

\begin{eqnarray}
\omega=a_0+a_1 x+a_2 x^2+a_3 x^3+a_4 x^4+a_5 x^5+a_6 x^6
\label{param}
\end{eqnarray}
where $x=|\mathbf{k}|/m_{\pi}$, and $a_0$, $a_1$, $a_2$, $a_3$, $a_4$, $a_5$, and $a_6$ are all expressed in GeV.\\

Here, the parametrization form of $\omega(\pi^0)$ at $I=0.3$ and $I=0.2$ is the same as in symmetric nuclear matter.

\begin{table*}[h]
\centering
{\small{\caption{The parameters for $\omega$ in symmetric nuclear matter.}
\begin{tabular}{c|c|c|c|c|c|c|c|c  }
  \hline
  \hline
Density                        &x               & $a_0$  & $a_1$ & $a_2$& $a_3$& $a_4$& $a_5$& $a_6$\\
  \hline
  \multirow{2}{*}{$0.5\rho_0 $} & $x\leq1.125$& 0.14762&0.00329&0.15947&-1.09393&3.2301&-3.73303&1.42962\\
  \cline{2-9}
                              &  $x>1.125$    &-0.08538&-0.99954&3.29955&-3.44383&1.71089&-0.41244&0.03884\\
  \hline
  \multirow{2}{*}{$\rho_0 $  } &$x\leq1.125$   & 0.16428&-0.10651&0.5233&-0.72545&0.28734&0&0\\
  \cline{2-9}
                               &$x>1.125$      &0.40054&-0.63405&0.50715&-0.15842&0.0181&0&0\\
\hline
  \multirow{2}{*}{$1.5\rho_0 $}&$x\leq1.125$   & 0.18147&0.04198&-0.53115&1.52268&-1.70649&0.62449&0\\
  \cline{2-9}
                               &$x>1.125$      &0.66757&-1.15148&0.83679&-0.25277&0.02837&0&0\\
\hline
  \multirow{2}{*}{$2\rho_0 $}&$x\leq1.125$   & 0.21177&-0.0107&-0.35514&0.96565&-1.09207&0.40349&0\\
  \cline{2-9}
                               &$x>1.125$      &0.83163&-1.50549&1.08558&-0.3289&0.03692&0&0\\
  \hline
  \hline
\end{tabular}}}
\label{tableI1}
\end{table*}

\begin{table*}[h]
\centering
{\small{\caption{The parameters for $\omega_{\pi^{+}}$  at isospin asymmetry $I=0.2$.}
\begin{tabular}{c|c|c|c|c|c|c|c|c  }
  \hline
  \hline
Density                        &x               & $a_0$  & $a_1$ & $a_2$& $a_3$& $a_4$& $a_5$& $a_6$\\
  \hline
  \multirow{2}{*}{$0.5\rho_0 $} & $x\leq1.125$& 0.14646&-0.01197&0.3134&-1.66811&4.26204&-4.60666&1.7103\\
  \cline{2-9}
                              &  $x>1.125$    &-0.07165&0.26725&-0.0744&0.00899&0        &0       &0\\
  \hline
  \multirow{2}{*}{$\rho_0 $  } &$x\leq1.125$   & 0.15305&0.14062&-1.03791&2.95229&-3.30208&1.23473&0\\
  \cline{2-9}
                               &$x>1.125$      &0.10479&-0.03411&0.08511&-0.02918&0.00353&0&0\\
\hline
  \multirow{2}{*}{$1.5\rho_0 $}&$x\leq1.125$   & 0.16982&0.09594&-0.78495&2.17423&-2.39857&0.88103&0\\
  \cline{2-9}
                               &$x>1.125$      &0.36904&-0.54203&0.41088&-0.1227&0.01366&0&0\\
\hline
  \multirow{2}{*}{$2\rho_0 $}&$x\leq1.125$   & 0.19358&0.04438&-0.57001&1.53095&-1.67771&0.61064&0\\
  \cline{2-9}
                               &$x>1.125$      &1.36471&-3.0588&2.86344&-1.30201&0.29304&-0.02607&0\\
  \hline
  \hline
\end{tabular}}}
\label{tableI2}
\end{table*}

\begin{table*}[h]
\centering
{\small{\caption{The parameters for $\omega_{\pi^{-}}$  at isospin asymmetry $I=0.2$.}
\begin{tabular}{c|c|c|c|c|c|c|c|c  }
  \hline
  \hline
Density                        &x               & $a_0$  & $a_1$ & $a_2$& $a_3$& $a_4$& $a_5$& $a_6$\\
  \hline
  \multirow{2}{*}{$0.5\rho_0 $} & $x\leq1.125$ & 0.14953&-0.0217&0.40913&-2.12781&5.18789&-5.4529&1.99392\\
  \cline{2-9}
                               &  $x>1.125$    &-0.03752&0.2044&-0.04916&0.00566&0        &0       &0\\
  \hline
  \multirow{2}{*}{$\rho_0 $  } &$x\leq1.125$   & 0.16601&-0.01132&0.14963&-0.30154&0.13559&0       &0\\
  \cline{2-9}
                               &$x>1.125$      &0.49951 &-0.77281&0.55235&-0.15885&0.01696 &0       &0\\
\hline
  \multirow{2}{*}{$1.5\rho_0 $}&$x\leq1.125$   & 0.18891&0.05333&-0.64292&1.713&-1.85345&0.6668 &0\\
  \cline{2-9}
                               &$x>1.125$      &0.86713&-1.568&1.12794&-0.34088&0.03816  &0       &0\\
\hline
  \multirow{2}{*}{$2\rho_0 $} &$x\leq1.125$   & 0.2235 &-2.27582&-0.53085&1.3709&-1.51562 &0.56282 &0\\
  \cline{2-9}
                               &$x>1.125$      &1.65668&-3.85346&3.62636&-1.65695&0.37366 &-0.03324&0\\
  \hline
  \hline
\end{tabular}}}
\label{tableI3}
\end{table*}

\begin{table*}[h]
\centering
{\small{\caption{The parameters for $\omega_{\pi^{+}}$ at isospin asymmetry $I=0.3$.}
\begin{tabular}{c|c|c|c|c|c|c|c|c  }
  \hline
  \hline
Density                        &x               & $a_0$  & $a_1$ & $a_2$& $a_3$& $a_4$& $a_5$& $a_6$\\
  \hline
  \multirow{2}{*}{$0.5\rho_0 $} & $x\leq1.125$& 0.14561&-0.01313&0.33908&-1.79495&4.55558&-4.90908&1.822\\
  \cline{2-9}
                              &  $x>1.125$    &-0.06849&0.25597&-0.06608&0.00771&0        &0       &0\\
  \hline
  \multirow{2}{*}{$\rho_0 $  } &$x\leq1.125$   & 0.15103&0.13677&-1.00479&2.89048&-3.2475&1.2163&0\\
  \cline{2-9}
                               &$x>1.125$      &0.02613&0.11939&-0.01774&0.00182&0       &0      &0\\
\hline
  \multirow{2}{*}{$1.5\rho_0 $}&$x\leq1.125$   & 0.19645&-0.07531&0.39858&-1.90247&4.05329&-3.87382&1.32371\\
  \cline{2-9}
                               &$x>1.125$      &1.9563&-4.52087&4.22417&-1.92159 &0.4317 &-0.03829  &0\\
\hline
  \multirow{2}{*}{$2\rho_0 $}&$x\leq1.125$     & 0.18406&0.07659&-0.71693&1.92654&-2.10031&0.76536&0\\
  \cline{2-9}
                               &$x>1.125$      &0.50864&-0.82079&0.60002&-0.17954&0.02003&0       &0\\
  \hline
  \hline
\end{tabular}}}
\label{tableI4}
\end{table*}

\begin{table*}[h]
\begin{center}
\small{\caption{The parameters for $\omega_{\pi^{-}}$ at isospin asymmetry $I=0.3$.}
\begin{tabular}{c|c|c|c|c|c|c|c|c  }
  \hline
  \hline
Density                        &x               & $a_0$  & $a_1$ & $a_2$& $a_3$& $a_4$& $a_5$& $a_6$\\
  \hline
  \multirow{2}{*}{$0.5\rho_0 $} & $x\leq1.125$& 0.14982&-0.00805&0.29261&-1.79699&4.812&-5.31034&1.9898\\
  \cline{2-9}
                              &  $x>1.125$    &-0.29222&0.68732&-0.38933&0.10963&-0.01166&0       &0\\
  \hline
  \multirow{2}{*}{$\rho_0 $  } &$x\leq1.125$   & 0.16437&0.09541&-0.77778&2.16077&-2.37432&0.86589&0\\
  \cline{2-9}
                               &$x>1.125$      &1.24039&-2.78125&2.6344&-1.20503&0.27254&-0.02435&0\\
\hline
  \multirow{2}{*}{$1.5\rho_0 $}&$x\leq1.125$   & 0.16901&-0.03369&0.38116&-1.92408&4.44758&-4.51472&1.61273\\
  \cline{2-9}
                               &$x>1.125$      &0.67832&-1.52577&1.60052&-0.79736&0.19558&-0.01883&0\\
\hline
  \multirow{2}{*}{$2\rho_0 $}&$x\leq1.125$   & 0.22697&0.0085&-0.62447&1.60553&-1.77879&0.66572&0\\
  \cline{2-9}
                               &$x>1.125$      &1.3686&-3.15732&2.95638&-1.34067&0.30092&-0.02673&0\\
  \hline
  \hline
\end{tabular}}
\label{tableI5}
\end{center}
\end{table*}

\end{appendices}

\end{document}